\documentclass[usenatbib]{mn2e}
\usepackage{graphicx, amssymb, aas_macros, bm, ulem, natbib}
\newcommand{\rmax}{R_{\rm{max}}}

\newcommand{\mvir}{M_{\rm{vir}}}

\newcommand{\rvir}{R_{\rm{vir}}}
\newcommand{\vvir}{V_{\rm{vir}}}

\newcommand{\mstar}{M_{\star}}
\newcommand{\msun}{M_{\odot}}

\newcommand{\hmpc}{h^{-1} \, {\rm Mpc}}

\newcommand{\mpc}{{\rm Mpc}}
\newcommand{\kpc}{{\rm kpc}}
\newcommand{\kms}{{\rm km \, s}^{-1}}
\newcommand{\millen}{MS-II}
\newcommand{\zacc}{z_{\rm acc}}

\newcommand{\macc}{M_{\rm acc}}

\newcommand{\lcdm}{$\Lambda$CDM}

\newcommand{\zfc}{z_{\rm fc}}
\newcommand{\tfc}{t_{\rm fc}}
\title[Magellanic Clouds in a \lcdm\ Universe]
{
Dynamics of the Magellanic Clouds in a \lcdm\ Universe
}
\author[M. Boylan-Kolchin, G. Besla, and L. Hernquist]{
  $\!\!$Michael~Boylan-Kolchin$^{1,2}$\thanks{$\!\!$e-mail: m.bk@uci.edu}, 
  Gurtina Besla$^3$, and Lars Hernquist$^3$\\
 $\!\!^1$Max-Planck-Institut f\"{u}r Astrophysik, Karl-Schwarzschild-Str. 1,
  85748 Garching, Germany\\
  $\!\!^2$Center for Galaxy Evolution,
    4129 Reines Hall, University of California, Irvine, CA 92697, USA\\
  $\!\!^3$Harvard-Smithsonian Center for Astrophysics, 60 Garden Street,
  Cambridge, MA 02138, USA}
\begin{document}

 \pagerange{\pageref{firstpage}--\pageref{lastpage}} 
 \pubyear{2010}

\maketitle

\label{firstpage}
\begin{abstract}
  We examine Milky Way-Magellanic Cloud systems selected from the Millennium-II
  Simulation in order to place the orbits of the Magellanic Clouds in a
  cosmological context.  Our analysis shows that satellites massive enough to be
  LMC analogs are typically accreted at late times.  Moreover, those that are
  accreted at early times and survive to the present have orbital properties
  that are discrepant with those observed for the LMC.  The high velocity of the
  LMC, coupled with the dearth of unbound orbits seen in the simulation, argues
  that the mass of the MW's halo is unlikely to be less than $2 \times
  10^{12}\,\msun$.  This conclusion is further supported by statistics of halos
  hosting satellites with masses, velocities, and separations comparable to
  those of the LMC.  We further show that: (1) LMC and SMC-mass objects are not
  particularly uncommon in MW-mass halos; (2) the apparently high angular
  momentum of the LMC is not cosmologically unusual; and (3) it is rare for a MW
  halo to host a LMC-SMC binary system at $z=0$, but high speed binary pairs
  accreted at late times are possible.  Based on these results, we conclude that
  the LMC was accreted within the past four Gyr and is currently making its
  first pericentric passage about the MW.
\end{abstract}

\section{Introduction} 

The Milky Way and its satellite galaxies offer a unique laboratory for
near-field cosmology.  The proximity of Milky Way (MW) satellites means that
their stellar content is resolved and can be used as a probe of galaxy formation
and evolution \citep{grebel2005}.  Furthermore, with the high astrometric
precision of instruments such as the Advanced Camera for Surveys on the Hubble
Space Telescope (HST), it is now also possible to measure accurate proper
motions for some of these satellites (e.g., \citealt{kallivayalil2006,
  kallivayalil2006a, piatek2007, piatek2008}).  These measurements have
significantly improved constraints on the satellites' orbital histories, and
have also revealed some surprises.  In particular, \citet[hereafter
K06]{kallivayalil2006} found that the velocity of the Large Magellanic Cloud
(LMC) is approximately $380\,\kms$, which is much larger than what typically had
been assumed ($\la 300 \,\kms$; e.g., \citealt{gardiner1996}) in modeling the
orbit of the LMC.

This large velocity has forced a reconsideration of the conventional picture of
the orbital history of the Magellanic Clouds (MCs), wherein the MCs have made
multiple passages about the MW over a Hubble time.  K06 and \citet[hereafter,
B07]{besla2007} analyzed possible LMC orbits and divided them into two
categories: early accretion, in which the LMC has made at least one complete
orbit about the Milky Way; and late accretion, where the LMC is currently making
its first pericentric passage.  Distinguishing between these two orbital
histories has important consequences for our understanding of the Local Group's
assembly history and for the formation of the Magellanic Stream
\citep{besla2010}.

The high velocities of the Clouds also raise the question of whether the MCs'
orbits are typical of massive satellites in MW-like systems and whether they can
provide information about the mass of the MW's halo.  B07 showed that the MCs
are effectively on an unbound orbit if the MW's mass is on the low end of
current estimates ($\sim 10^{12}\,\msun$, consistent with the results of
\citealt{xue2008} based on blue horizontal-branch stars).  On the other hand, a
massive MW halo ($\sim [2-2.5] \times 10^{12}\msun$) -- in line with
estimates based on the timing argument and satellite kinematics including data
from Leo I \citep{li2008a, watkins2010} -- implies that the LMC has a velocity
that is substantially lower than the local escape speed.

Analyses of cosmological simulations indicate that unbound orbits are quite rare
\citep{van-den-bosch1999, vitvitska2002, benson2005, khochfar2006, diemand2007a,
  wetzel2010a}: for example, \citet{wetzel2010a} found that less than 2\% of
merging satellites have formally unbound orbits at all masses and redshifts.
\citet{sales2007} have shown that a non-negligible number of subhalos in
cosmological simulations can be scattered to high energy orbits as the result of
three-body encounters, but these subhalos are dynamically required to be
low-mass, likely rendering this mechanism irrelevant for the LMC.  Almost all of
these results are based on modeling dark matter halos as point mass, Kepler
potentials.  This is a poor approximation to the true structure of dark matter
halos at small radii, however. The LMC, which is currently at $\la$ one-fifth
the MW's virial radius, is in precisely this regime.  As a result, it is far
from clear whether an unbound LMC would be highly unusual in a cosmological
context.

The very existence of massive satellites such as the MCs can also be used to
place constraints on the mass of the MW.  If the LMC is typical for galaxies of
its stellar mass, it likely had a dark matter mass of $[1-2] \times 10^{11}
\,\msun$ before accretion by the MW \citep{guo2010, boylan-kolchin2010}.  The
SMC has a stellar mass that is typical for halos that are only a factor of 2-3
less massive than this.  \citet[hereafter BK10]{boylan-kolchin2010} used
statistics of subhalos from a large sample of simulated MW-mass halos to
conclude that MC-mass galaxies are rare if the MW has $M \la 10^{12}\,\msun$ but
are much more typical if the MW has a massive dark matter halo ($\sim 2.5 \times
10^{12}$).  These results were based purely on the masses of the MCs and MW,
however, and did not include the additional constraints provided by the MCs'
orbits.

In this paper, we examine the orbits, accretion epochs, and masses of MW-MC
systems using the Millennium-II Simulation (MS-II;
\citealt{boylan-kolchin2009}), which has sufficient mass resolution to probe
SMC-mass scales and a large enough volume to contain a large, statistical sample
of MW-mass halos.  We use realistic -- but still spherically-symmetric --
potentials that are expected for LCDM dark matter halos when computing orbital
parameters in an attempt to accurately describe the orbits of MC analogs at
small radii from their hosts.

Our goal is to investigate five main questions related to the MCs in order to place their
orbital properties and masses in a cosmological context:
\begin{itemize}
\item How common are satellites with masses similar to those of the LMC and SMC
  within Milky Way-mass halos at redshift zero?
\item How typical is the LMC in terms of its orbital properties (e.g., energy
  and angular momentum)?
\item Are LMC-type satellites at $z=0$ typically accreted {\bf early} (having
  completed at least one pericentric passage about their hosts) or {\bf late}
  (being on their first infall towards their hosts)?
\item How likely is it that the MCs were accreted as a binary system?
\item What can we infer about host halos from the properties of massive
  satellites such as the LMC?
\end{itemize}

Our work is structured as follows.  \S~\ref{section:methods} provides relevant
details about the \millen, describes our assumptions about the MW and MCs, and
defines our subhalo samples.  In \S~\ref{subsection:FirstCross}, we determine
the most likely accretion epoch for the LMC based on mass considerations.  In
\S~\ref{subsection:ang}-\S~\ref{subsec:eccentricity}, we fold the orbital
properties of the LMC into the analysis to distinguish between the early or late
accretion scenarios.  Section~\ref{section:Mass} explores the expected frequency
of L/SMC-type companions about MW-type hosts based on the MCs' expected infall
masses.  In \S~\ref{section:LS}, we examine potential L/SMC binary systems and
their properties.  Our results are discussed in \S~\ref{section:Discuss}; in
particular, we assess the likelihood that the MCs are on their first passage
about the MW (\S~\ref{subsection:First}).  We present our conclusions in
\S~\ref{section:Conclusions}.

\section{Methodology}
\label{section:methods}
\subsection{The Millennium-II Simulation}
\label{subsec:msII}
Our analysis of objects similar to the Magellanic Clouds is based on the
Millennium-II Simulation, a very large $N$-body simulation that follows the
evolution of over ten billion particles in a periodic cube of $(137\,\mpc)^3$
from redshift 127 to 0.  The cosmological parameters used in the \millen\ are
identical to those adopted for the Millennium Simulation \citep{springel2005b}
and the Aquarius simulations \citep{springel2008}:
\begin{eqnarray}
 \label{eq:cosmo_params}
 & & \Omega_{\rm tot} = \! 1.0, \; \Omega_m = \!0.25, \; \Omega_b=0.045, \; 
 \Omega_{\Lambda}=0.75, \nonumber \\
 & & h = 0.73, \;\sigma_8=0.9, \; n_s=1\,,
\end{eqnarray}
where $h$ is the Hubble constant at redshift zero in units of $100 \, \kms\,
{\rm Mpc}^{-1}$, $\sigma_8$ is the rms amplitude of linear mass fluctuations in
$8 \,\hmpc$ spheres at $z=0$, and $n_s$ is the spectral index of the primordial
power spectrum.  The \millen\ offers a unique combination of mass resolution and
large volume for studying dynamics of Magellanic Cloud analogs within Milky
Way-mass systems: the \millen\ particle mass of $m_p=9.43 \times 10^6\,\msun$
results in over 100,000 particles in Milky Way-mass halos at $z=0$, and resolves
LMC-mass subhalos with $\ga 10,000$ particles.  For further information about
the \millen, see \citet{boylan-kolchin2009}.\footnote{Merger trees, along with
  halo and subhalo catalogs, from the \millen\ are publicly available at
  http://www.mpa-garching.mpg.de/galform/millennium-II/ .}

\subsection{Milky Way-mass halos}
\label{subsec:MW}
Current estimates of the mass of the Milky Way's halo range from $\approx [1-3]
\times 10^{12}\,\msun$ (e.g., \citealt{sakamoto2003, battaglia2005, dehnen2006,
  xue2008, li2008a, gnedin2010, watkins2010}; and references therein).  In order
to bracket this range, and to understand any trends with the mass of the MW, we
select all halos with $4.3 \times 10^{11} \le \mvir/\msun \le 4.3 \times
10^{12}$ from the \millen\ at redshift zero as our primary ``Milky Way'' sample.
This set is identical to that of \citet{boylan-kolchin2010} and contains
approximately 7600 dark matter halos, 2658 of which have $\mvir \in [1-3] \times
10^{12}\,\msun$.  It is important to note that not all of these halos are
expected to host MW-like galaxies in the standard \lcdm\ model: for exmple,
approximately 30\% of halos in our full MW sample should host galaxies that are
not late-type based on their colors and specific star formation rates
\citep{weinmann2006}.  By taking subsets of our main sample that are based on,
e.g., environment, we can explore whether host halo properties correlate with
the likelihood of hosting satellite galaxies like the MCs.

We use the radius inside of which the average density exceeds the critical
density of the universe by a factor $\Delta_{\rm vir}$ (see \citealt{eke1996}
and \citealt{bryan1998} for details) as the virial radius $\rvir$ of our halos,
and the mass enclosed within this radius, $\mvir$, as the virial mass.  At
redshift zero in the cosmology of the Millennium-II Simulation, $\Delta_{\rm
  vir}=94.2$ and the virial radius $\rvir$ and virial (circular) velocity
$\vvir$ scale with the virial mass as:
\begin{eqnarray}
  \label{eq:vvir}
  \rvir &=&257.8 \,\left(\frac{\mvir}{10^{12} \,\msun}\right)^{1/3} \kpc \\
  \vvir &=& 129.2 \,\left(\frac{\mvir}{10^{12} \,\msun}\right)^{1/3} \kms \,.
\end{eqnarray}
We will use these relations in subsequent sections to scale measured properties
of the Magellanic Clouds -- e.g., their angular momenta -- to the virial values
for a range of virial masses for the MW.

Later sections will also focus on orbital properties of subhalos within their
hosts.  These properties will be computed assuming that the host dark matter
halos are well-fitted by Navarro, Frenk, and White (\citeyear{navarro1996},
\citeyear{navarro1997}; hereafter, NFW) density profiles.  The structure of the
NFW profile is fully specified by a halo's mass and the radius at which the
circular velocity curve peaks, $\rmax$.  Using values of $\mvir$ and $\rmax$
computed directly from the \millen\ to set the NFW potential, we calculate the
energy, angular momentum, and eccentricity of orbiting subhalos within MW-mass
hosts\footnote{We assume a fixed concentration of 10 when computing the
  potential of the actual MW.  This is not a strong assumption, as changing the
  concentration between, e.g., 8 \& 12 does not significantly affect the
  inferred pericenter, angular momentum, or orbital energy of the subhalos.}.
We defer an analysis of the current distance of the LMC from the MW to
  Section~\ref{subsec:bayes}, as it is inherently extremely rare for a satellite
  to be near pericenter (as is the case for LMC) for the eccentric orbits
  typical of \lcdm\ satellites \citep{tormen1997, diemand2007a}.
Note that one limitation of our approach is the use of spherically symmetric
profiles in our calculations, whereas halos from cosmological dark matter
simulations are typically prolate or triaxial (e.g., \citealt{warren1992,
  jing2002, allgood2006, bett2007}).

\subsection{The Magellanic Clouds}
\label{subsec:MC}
\subsubsection{Observed properties}
\label{subsubsec:MC_obs}
The Large (Small) Magellanic Cloud is the most (second-most) luminous satellite
of the Milky Way.  Following \citet{kim1998}, we assume that the stellar mass of
the LMC is $M_{\rm LMC, \star}=2.5 \times 10^{9}\,\msun$.  We adopt a
galactocentric distance of $R_{\rm LMC}=50.1 \,\kpc$ \citep{freedman2001}.
K06's analysis of the LMC's proper motion shows that
\begin{eqnarray}
  \label{eq:lmc_v}
  V_{\rm LMC, tan} \! &=&\! 367 \pm 18 \;\kms \nonumber\\
  V_{\rm LMC, rad} \! &=&\! \;\; 89 \pm 4 \;\kms\,.
\end{eqnarray} 
The specific angular momentum of the LMC, normalized by the specific angular
momentum of a circular orbit at the MW's virial radius, is therefore
\begin{equation}
  \label{eq:jlmc}
  \widetilde{j}_{\rm LMC} \equiv \frac{R_{\rm LMC} \,V_{\rm LMC,tan}}{\rvir
    \,\vvir} = 0.55 
  \left(\frac{\mvir}{10^{12} \,\msun}\right)^{-2/3} \,.
\end{equation}

We assume that the SMC is at a distance of 58.9 kpc \citep{kallivayalil2006a}
and has a stellar mass of $M_{\rm SMC, \star}=3 \times 10^{8}\,\msun$
\citep{stanimirovic2004}.  The proper motion measurements of
\citet{kallivayalil2006a} give
\begin{eqnarray}
  \label{eq:smc_v}
  V_{\rm SMC, tan} \! &=&\! 301 \pm 52 \;\kms \nonumber\\
  V_{\rm SMC, rad} \! &=&\! \;\; 23 \pm 7 \;\kms\,.
\end{eqnarray} 
From these values, the specific angular momentum of the SMC is
\begin{equation}
  \label{eq:jsmc}
  \widetilde{j}_{\rm SMC} \equiv \frac{R_{\rm SMC} \,V_{\rm SMC,tan}}{\rvir \,\vvir} = 0.53
  \left(\frac{\mvir}{10^{12} \,\msun}\right)^{-2/3} \,.
\end{equation}
The angular momenta of the LMC and SMC are therefore strikingly close to each
other.

\citet{piatek2008} have performed an independent analysis of the HST proper
motion data and find results that are in general agreement with, but at the
lower end of, the range found by K06.  In particular, Piatek et al. find
($V_{\rm tan}$, $V_{\rm rad}$) of ($346 \pm 8.5\,\kms$, $93.2 \pm 3.7\,\kms$)
and ($259 \pm 17\,\kms$, $6.8 \pm 2.4\,\kms$) for the Large and Small MCs,
respectively.  Adopting these values changes the normalization of
Eqn.~(\ref{eq:jlmc}) to 0.52 and of Eqn.~(\ref{eq:jsmc}) to 0.46.  Recently,
\citet{vieira2010} determined proper motions for the MCs using photographic and
CCD observations from the Yale/San Juan Southern Proper Motion program spanning
a baseline of 40 years.  They also confirm the K06 results and find proper
motions for the SMC that are more consistent with the \citet{kallivayalil2006a}
analysis than that of Piatek et al.

\subsubsection{Magellanic Clouds in the Millennium-II Simulation}
\label{subsubsec:MC_sim}
In order to identify $z=0$ analogs of the LMC, we first search our full MW
sample for all subhalos that survive to $z=0$ and reside within a fixed fraction
of $\rvir$ of their host (see below). For each of these subhalos, we also find
the mass of its main progenitor at all earlier times.  To define analogs of the
Magellanic Clouds in an $N$-body simulation such as the MS-II, we need a way to
link dark matter (sub)halos to galaxies.  Recent work has shown that many
observed statistical properties of galaxies can be reproduced under the simple
``abundance matching'' assumption that stellar mass is a monotonically
increasing function of the maximum dark matter mass a subhalo attains over its
history (e.g., \citealt{conroy2006, wang2006, moster2010, guo2010, klypin2010}).
Accordingly, we select as our fiducial sample the subhalo with the largest
maximum main progenitor mass in each halo; we will refer to this sample as the
first-ranked subhalos.  The mass of this progenitor and the redshift at which it
was reached are denoted $\macc$ and $\zacc$; we will sometimes refer to these
quantities as the infall mass and infall redshift.  Note that subhalos selected
in this manner are not necessarily the most massive subhalos in their hosts at
$z=0$.

We identify LMC analogs within this sample by adding two additional conditions.
First, the host halos must obey $\mvir \in [1-3] \times 10^{12}\,\msun$,
corresponding to approximately 35\% of our full host halo sample; note that this
is in the upper end of our host sample in terms of mass.  Next, we use abundance
matching as implemented by \citet{guo2010} to estimate the maximum dark matter
halo mass of the LMC.  With our definition of halo mass, the infall halo mass
corresponding to the LMC's stellar mass is $M_{\rm LMC}(\zacc) \approx 1.6
\times 10^{11}\,\msun$.  We define LMC analogs as first-ranked subhalos having
infall masses within a factor of two of this mass, $8 < M_{\rm LMC}(\zacc) \,
[10^{10}\,\msun] < 32$, which allows for both scatter in the $\mstar-M_{\rm
  halo}$ relation and uncertainty in the stellar mass of the LMC.  Such subhalos
are more massive than the typical most massive subhalo, in terms of maximum
progenitor mass, of MW-mass halos (BK10; see \S\ref{section:Mass} for a full
analysis).

To identify SMC analogs, we first find the surviving subhalos at $z=0$ having
the second highest infall mass among all subhalos of their hosts; we will refer
to this fiducial sample as the second-ranked subhalos.  Using the SMC's stellar
mass, we select second-ranked subhalos with masses between $4 \times 10^{10}$
and $1.6 \times 10^{11} \, \msun$ in hosts with $\mvir \in [1-3] \times
10^{12}\,\msun$ as our SMC analogs.  All SMC analogs therefore have at least
4000 particles at their maximum, well above the $\sim 1500$ particle requirement
that \citet{guo2010} show is necessary to adequately resolve the $\macc$ halo +
subhalo mass function at $z=0$ (see also \citealt{wetzel2010}).

It is important to choose a suitable limiting radius for defining each halo's
most (second-most) massive surviving subhalo.  A natural choice is the redshift
zero virial radius.  We therefore consider the most (second-most) massive
  non-dominant subhalo within $\rvir$ at $z=0$ as our fiducial sample.  The
  precise choice of limiting radius used is not particularly crucial, so long as
  the radius is sufficiently large to cover much of the halo: we find similar
  results for $0.65-1.0\,\rvir$.

To summarize, we define two samples for both the LMC and the SMC:
\begin{itemize}
\setlength{\itemindent}{1em}
\item {\bf first-ranked subhalos} (1st): \hfill \\
  the most massive subhalo, in terms of the maximum mass ever attained, located
  within $\rvir$ of a host's center at $z=0$.  This sample
  contains 7641
  subhalos.
\item {\bf LMC analogs}: \\
  the subset of {\bf 1st} with $8 \times 10^{10} < \macc/\msun <3.2 \times
  10^{11}$ located in hosts with virial mass $\mvir \in [1-3] \times 10^{12}\,\msun$.
  This sample contains {\bf 938} subhalos.  
\item {\bf second-ranked subhalos} (2nd): \\
  the second most massive subhalo, in terms of the maximum mass ever attained,
  located within $\rvir$ of a host's center at $z=0$.  This sample
  contains 7639 subhalos.
\item {\bf SMC analogs}: \\
  the subset of {\bf 2nd} with $4 \times 10^{10} <  \macc/\msun < 1.6 \times
  10^{11}$ located in hosts with virial mass $\mvir \in [1-3] \times 10^{12}\,\msun$.
  This sample contains {\bf 840} subhalos.
\end{itemize}

\section{Orbital properties of LMC candidates}
\label{section:orbits}
\subsection{First crossing time}
\label{subsection:FirstCross}
\begin{figure}
 \centering
 \includegraphics[scale=0.55, viewport=0 0 410 440]{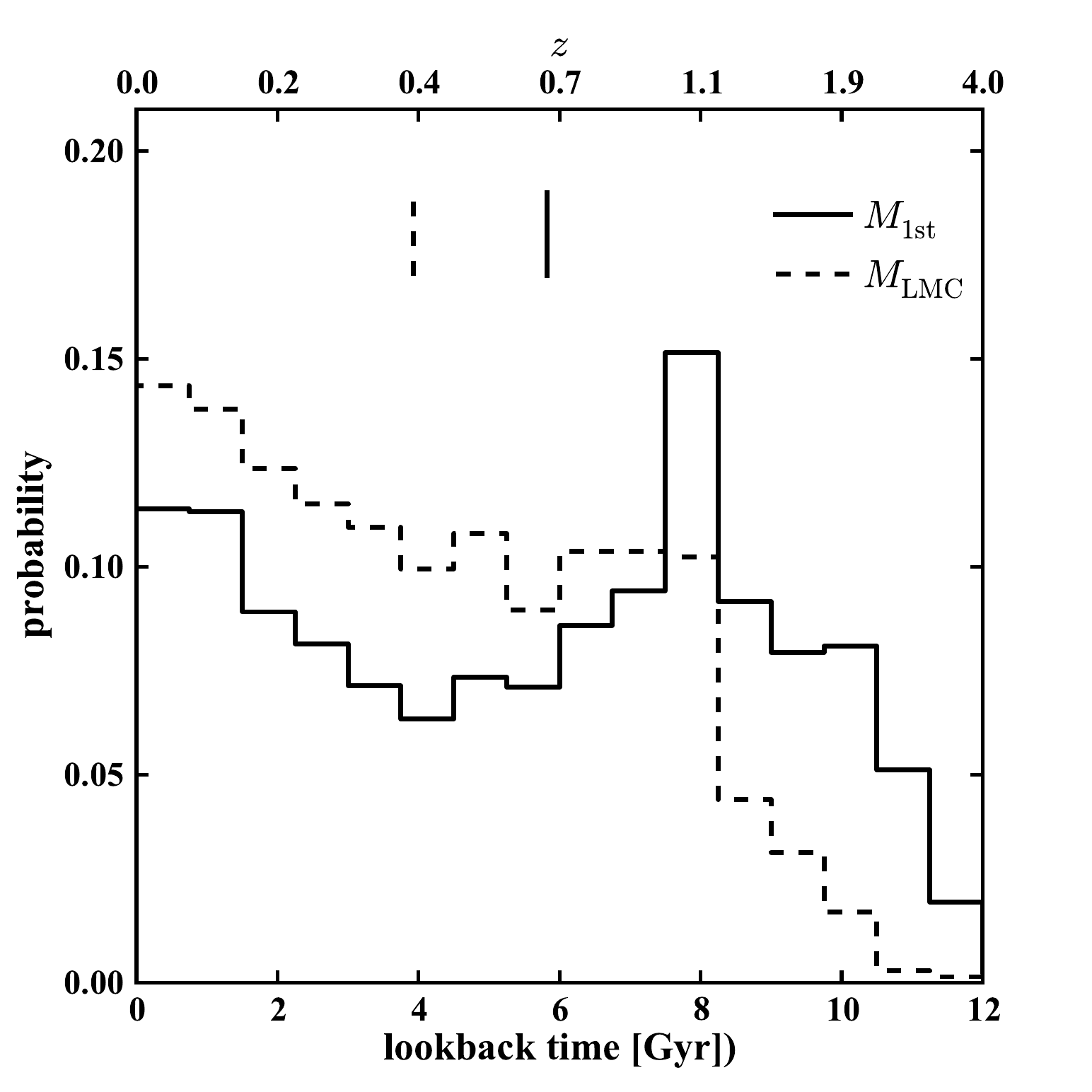}
 \includegraphics[scale=0.55, viewport=0 0 410 440]{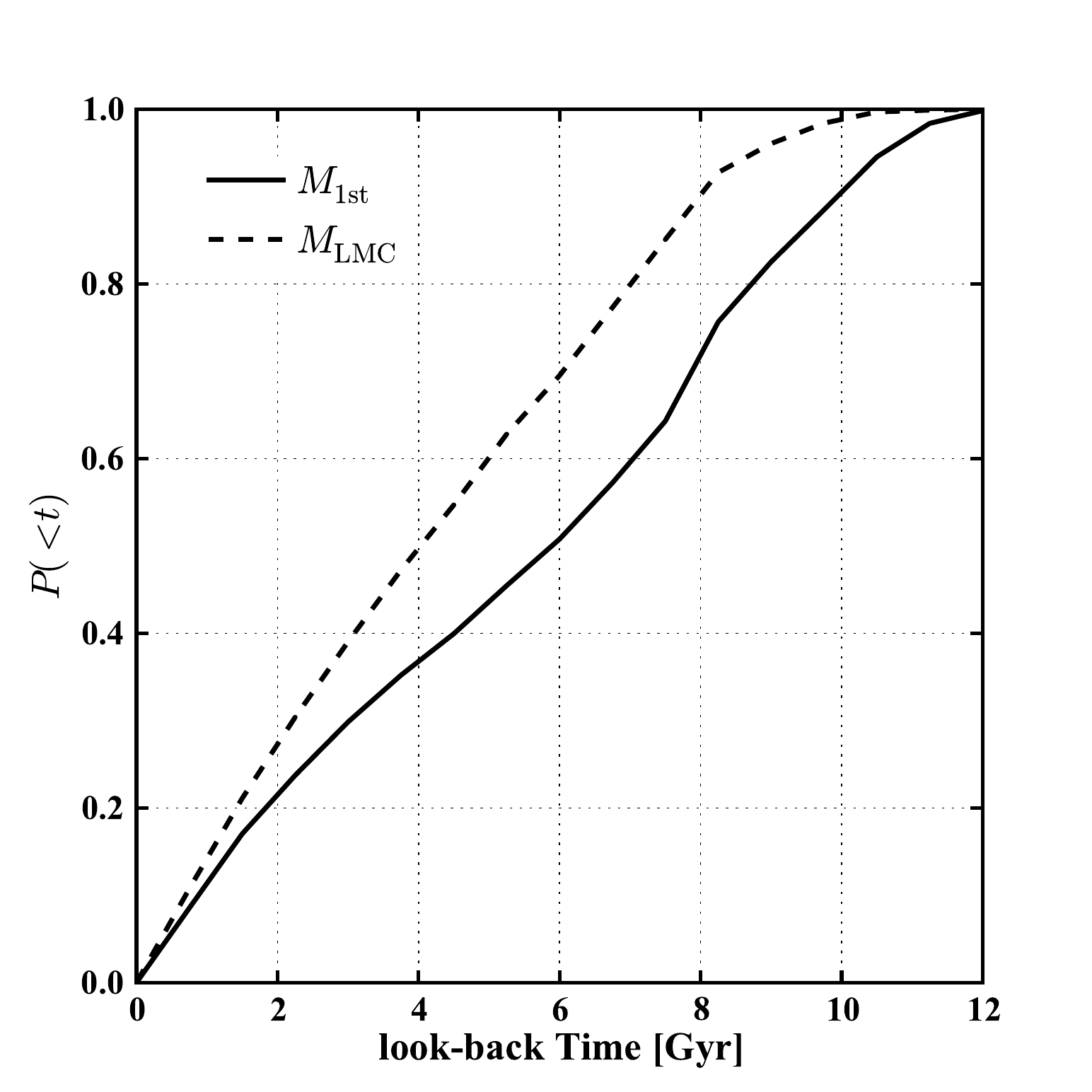}
 \caption{{\it Top:} Lookback time when the 1st ranked subhalo (solid
   histogram) or LMC analog (dashed histogram) first crossed the physical $z=0$ virial
   radius of its host.  Vertical lines mark the medians of the distributions.
   {\it Bottom:}  Cumulative version of the upper plot. 50\% of LMC analogs were accreted
   by MW-mass hosts within the past 4 Gyr and 25\% within the past 2 Gyr. 
 \label{fig:firstCrossDist}
}
\end{figure}

The lookback time at which the LMC first crossed the physical $z=0$ virial
radius of the MW, moving inward -- referred to hereafter as the first crossing
time ($\tfc$) -- serves as a useful discriminant between the early and late
accretion scenarios laid out in B07.  We therefore compute $\tfc$ for all
first-ranked halos in our full MW sample to investigate whether there is a
preferred accretion epoch for LMC-like objects.

The distributions of $\tfc$ for all first-ranked subhalos and for those in the
LMC analog subsample are shown in Fig.~\ref{fig:firstCrossDist} as solid and
dashed histograms, respectively.  The full distribution is bimodal, with a
primary peak at $t \approx 8$ Gyr and a secondary peak at $t=0$ Gyr; the median
value is $\tfc=5.8$ Gyr.  The distribution for the LMC-mass sample is markedly
different.  There is no prominent peak at $\tfc \approx 8$ Gyr; instead, the
distribution rises continuously from large lookback times (high redshift) to the
present day, and the median of the distribution lies at $\tfc = 3.9$ Gyr.  The
differences between the two distributions reflect the relatively high masses of
the LMC analog sample: since these subhalos are more massive than the average
first-ranked subhalo, they are accreted onto their host halos later, in the
typical hierarchical manner expected in the \lcdm\ cosmology.

Candidates in the LMC's expected mass range are usually accreted at fairly late
times: only 12\% have $\tfc > 7.5$ Gyr ($\zfc > 1$), while approximately 30\%
have been accreted within the past 2 Gyr.  Such numbers favor the late accretion
scenario for the LMC.  Even taking all first-ranked subhalos (the solid line in
the lower panel of Fig.~\ref{fig:firstCrossDist}), we find that $\sim$ 70\% were
accreted since $z=1$.  First-ranked subhalos that could have completed several
orbits are rare.

\begin{figure}
 \centering
 \includegraphics[scale=0.55, viewport=0 0 410 440]{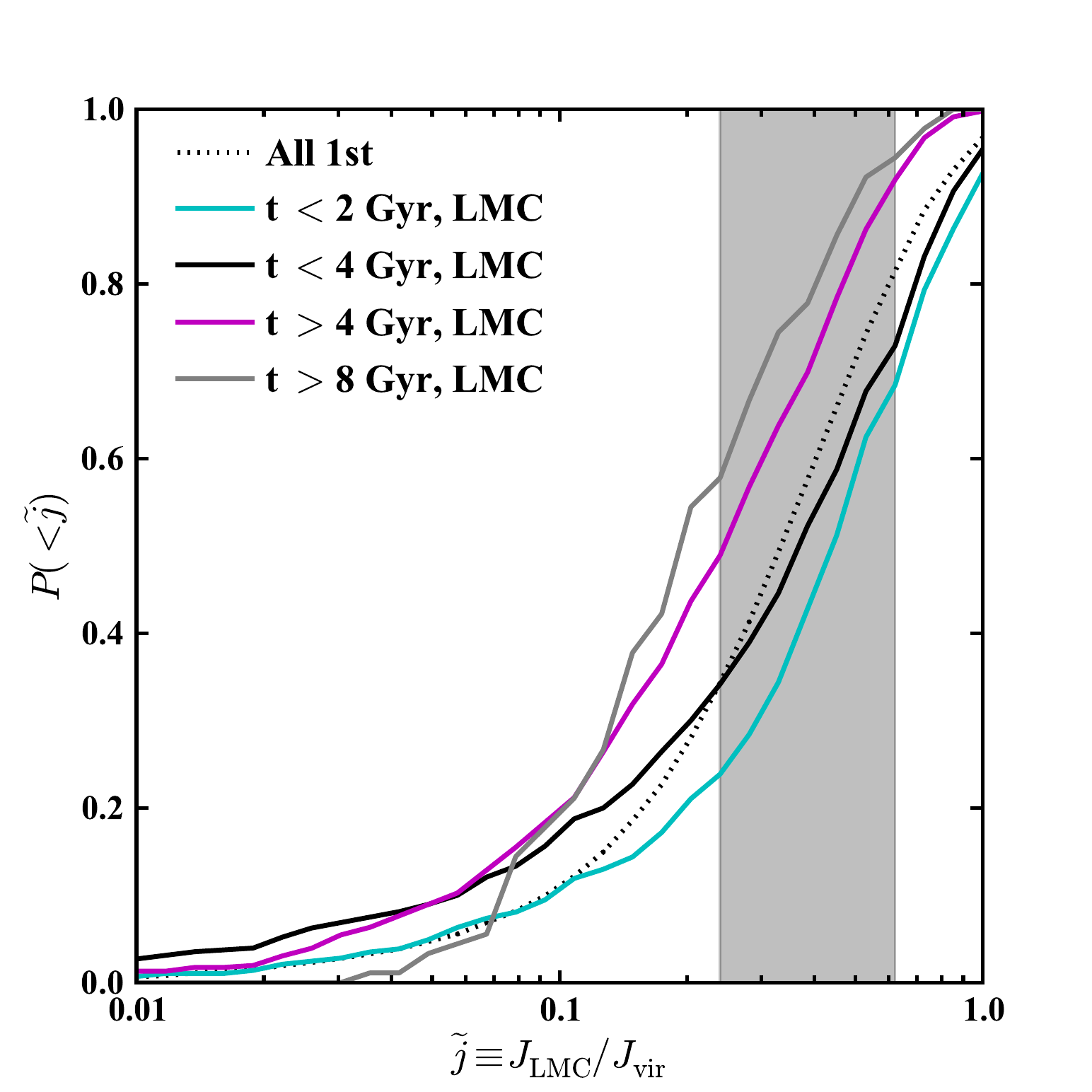}
 \caption{
  Distribution of (specific) angular momentum for LMC analogs
   accreted early (more than 4 
   Gyr ago; magenta) and late (within the last 4 Gyr; black), as well as very
   early (more than 8 Gyr ago; gray) and very late (less than 2 Gyr ago;
   cyan).  Also plotted is the cumulative distribution for all 1st-ranked
   subhalos, independent of accretion epoch (black dotted line).
   The gray 
   shaded region
   corresponds to the observed angular momentum of the LMC, including $\pm
   2\sigma$ errors on the tangential velocity, for
   a Milky Way mass in the range $[1-3]\times 10^{12}\,\msun$.  Late-accreted
   LMCs typically have higher specific angular momentum, due to both the
 inefficiency of dynamical friction over $\sim 3$ Gyr time-scales and the higher
 specific angular momentum of late-accreted subhalos.
  \label{fig:j_jvir}
}
\end{figure}

\subsection{Angular momentum}
\label{subsection:ang}

Figure~\ref{fig:j_jvir} shows the cumulative distribution of specific angular
momenta\footnote{The angular momentum is computed with respect to the host's
  center, defined by the location of the gravitational potential minimum.  The
  host's velocity is determined by averaging over all particles in the host's
  main subhalo.  We have also tried computing the velocity with respect to only
  the most bound particles in the host subhalo or only those particles within 10
  or 25 kpc and found that the results presented here are unchanged.}
$j=R\,V_{\rm tan}$, normalized by the virial value $j_{\rm vir}=\rvir \,\vvir$,
for the full 1st-ranked subhalo sample (dotted line).  Also plotted is the same
quantity for LMC analogs, split based on their accretion times: less than 2 and
4 Gyr ago (cyan and black lines, respectively), and more than 4 and 8 Gyr ago
(magenta and gray lines).  The gray shaded region marks the range of
$\widetilde{j}=j/j_{\rm vir}$ allowed for the LMC based on a MW mass of $10^{12}
\le \mvir \le 3\times 10^{12}\,\msun$ and including $\pm 2 \sigma$ on the LMC's
tangential velocity.  This allowed range is fairly broad and shows that the
angular momentum of the observed LMC is, in fact, fairly typical of LMC analogs
in the \millen.

Comparing the angular momentum of LMC analogs based on accretion epoch, we find
that LMCs accreted $>$ 4 Gyr ago have much lower angular momentum than those
accreted at later times. In fact, 50\% of those LMCs accreted $>$4 Gyr ago have
angular momentum lower than the observed range for the LMC.  On the other hand,
half of the LMC analogs accreted within the past 2 Gyr have angular momentum
within the shaded region, supporting a late accretion scenario for the LMC.

\begin{figure}
 \centering
 \includegraphics[scale=0.55, viewport=0 0 410 440]{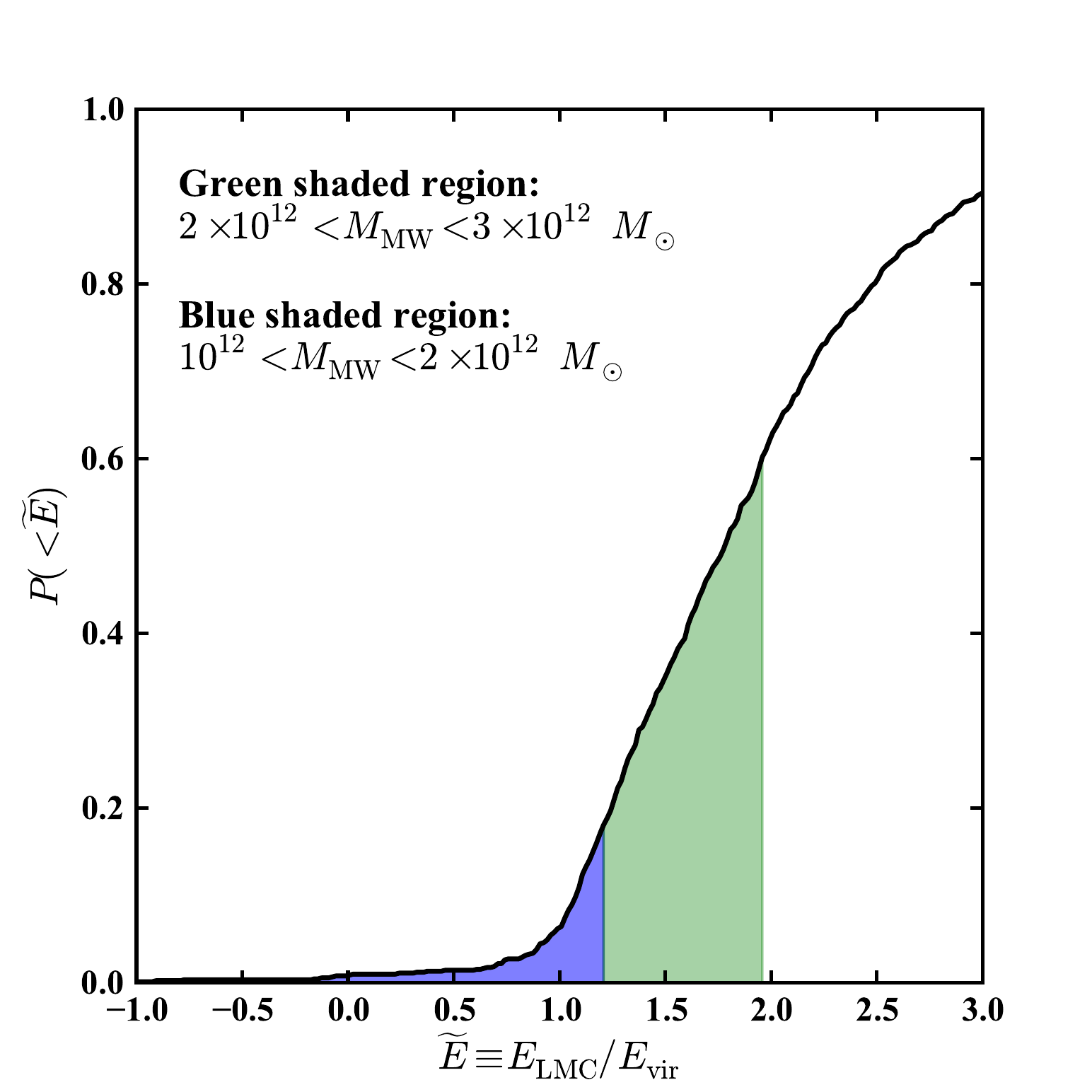}
 \caption{Distribution of orbital energies for the LMC analogs relative to the
   energy of a circular orbit at the host's virial radius.  
   Blue and green shaded
   regions correspond to hosts of $[1-2]\times 10^{12}$ and $[2-3]\times
   10^{12}\,\msun$, respectively.  Unbound orbits ($\widetilde{E} < 0$) are
   extremely rare, as are orbits consistent with a low-mass ($\la
   1.5 \times 10^{12}\,\msun$) MW.
 \label{fig:e_cum}
}
\end{figure}
Early accreted ($>$ 4 Gyr ago) LMC analogs in low-mass MWs are strongly
disfavored:  even in the most extreme scenario of $\mvir=3\times10^{12}$ and
  $V_{\rm tan, LMC}=331\,\kms$ ($2\,\sigma$ lower than the central value of
  K06), we still find that over 50\% of LMC analogs have a lower value of
  $\widetilde{j}$ than for the actual LMC.  A more realisitc scenario --
$\mvir=1.6 \times 10^{12}$ and $V_{\rm tan, LMC}=367\,\kms$ -- results 90\% of
early-accreted LMCs having lower angular momentum than is observed at $z=0$.

Regardless of accretion epoch, approximately 30-35\% of LMC analogs fall in the
gray shaded region: the LMC's angular momentum is not atypical in a cosmological
context.  This result dismisses previous assertions that tidal torques from M31
are needed to explain the orbital angular momentum of the LMC (e.g.,
\citealt{raychaudhury1989, shuter1992, byrd1994}).  Originally, concerns about the
LMC's angular momentum arose because the orbital plane of the MCs is polar
to the disk plane of the MW, while the LMC has orbital angular momentum that is
at least as much as that of the MW's thin disk \citep{fich1991, lin1995,
  sawa2005}.  This is potentially difficult to explain in an early accretion
scenario: torques from the MW's disk certainly could not explain the angular
momentum of the LMC in this configuration, leading to the search for alternate
potential perturbers.
While our analysis does not extend to the likelihood of a polar orbital
orientation, we do find that the angular momentum of the LMC's orbit is more
typical in a recent accretion scenario. Over such short interaction timescales,
torques from the host are largely irrelevant and our result should hold
regardless of the orientation of the orbit.

\subsection{Energy}
\label{subsection:energy}
\begin{figure}
 \centering
 \includegraphics[scale=0.55, viewport=0 0 410 440]{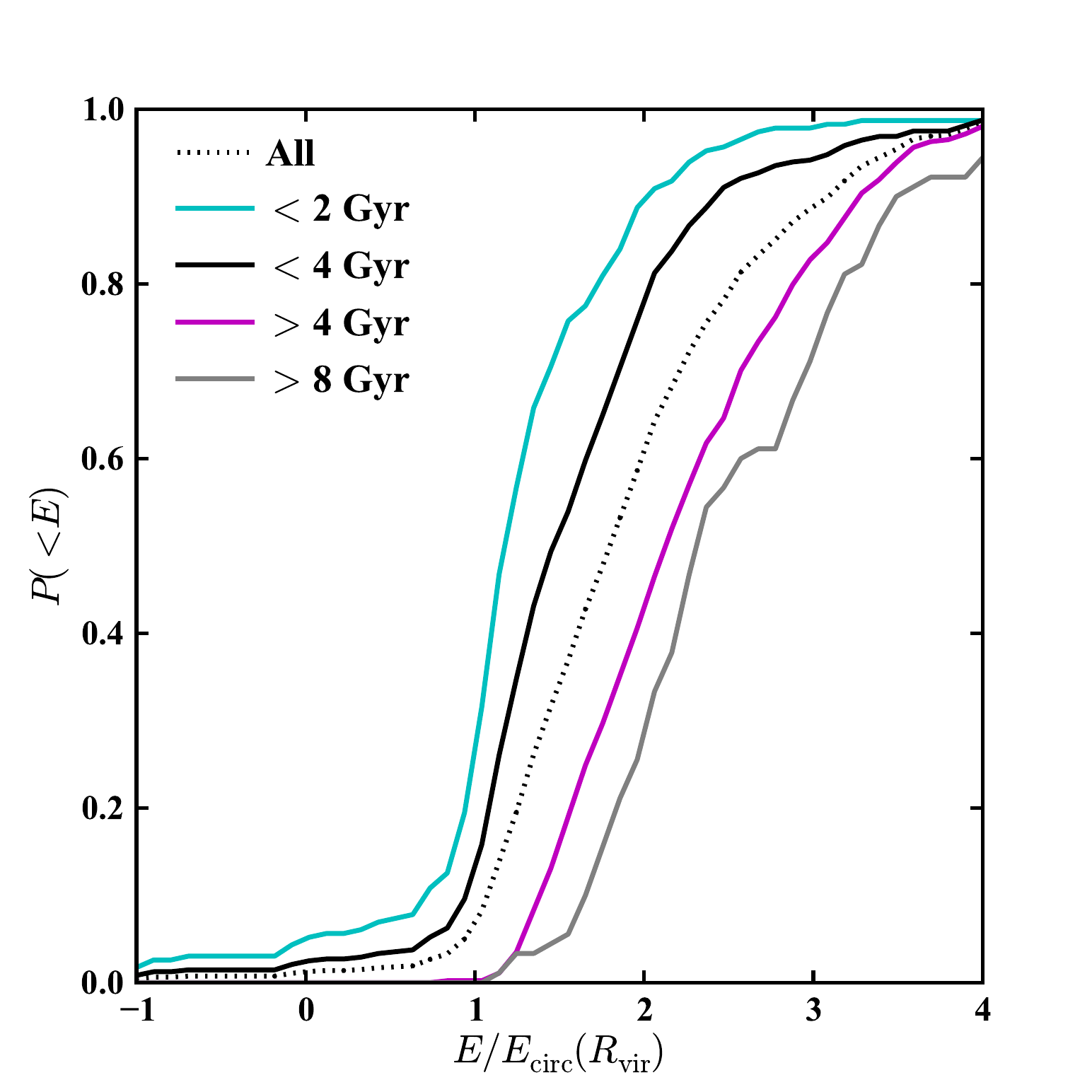}
 \caption{Orbital energies for the LMC analog sample (dotted curve), as well as 
   results split by $\tfc$ (solid curves).  The early-accreted LMCs ($>$ 4 Gyr,
   shown in magenta) are typically more bound to their host than are the 
   late-accreted LMCs ($<$ 4 Gyr, shown in black).  This trend is even stronger
   for LMCs accreted over 8 Gyr ago (gray) versus those accreted within the past
   2 Gyr (cyan).
  \label{fig:e_sham}
}
\end{figure}

Figure~\ref{fig:e_cum} shows the orbital energy distribution of the \millen\ LMC
analogs, normalized by the energy of a circular orbit at the host's virial
radius.  A striking feature of Fig.~\ref{fig:e_cum} is that nearly all subhalos
are on bound orbits ($\widetilde{E} \equiv E/E_{\rm circ}(R_{\rm vir}) > 0$).
The implications of the energies of subhalos' orbits for the LMC are highly
sensitive to the virial mass of the MW: less than 10\% of LMC analogs have
orbits as energetic as that of the observed LMC if the MW's virial mass is
smaller than $2\times 10^{12}\,\msun$ (blue shaded region), while a substantial
fraction have orbits that match the energy of the LMC's orbit if the MW's halo
lies between $[2-3]\times 10^{12}\,\msun$ (green shaded region).  Even if we
take both radial and tangential velocities that are $1\,\sigma$ ($2\,\sigma$)
lower than K06's mean values, we still find that only 2.3\% (7.9\%) of orbits
are more energetic than observed for the LMC in a halo of $1.5 \times
10^{12}\,\msun$.

In Figure~\ref{fig:e_sham} we show the cumulative distribution of $\widetilde{E}
$ for the LMC analog sample, split by accretion epoch (solid curves); the
distribution for the full LMC analog sample (Fig.~\ref{fig:e_cum}) is shown as
the dotted curve.  There is a marked difference when looking at early ($> $4
Gyr) versus late ($<$ 4 Gyr) accreted LMCs.  Early-accreted LMCs tend to be on
much more bound orbits (higher values of $\widetilde{E}$), while late-accreted
LMCs are less bound to their host halos (though virtually all are still bound,
formally).

Using the mean velocities from K06, we find that a Milky Way of mass
$(1,2,3)\times 10^{12}\,\msun$ corresponds to $\widetilde{E}=(-0.53, 1.21,
1.96)$.  If the Milky Way's halo mass does not exceed $2 \times 10^{12}\,\msun$,
the early accretion scenario is strongly disfavored: there are vanishingly few
early-accreted LMCs in the \millen\ having $\widetilde{E} < 1.2$.
Fig.~\ref{fig:e_sham} also strongly disfavors {\it any} LMC accretion scenario
for a $10^{12}\,\msun$ MW (which puts the LMC on an unbound
orbit). Approximately 25\% of the late-accreted (and 7\% of the early-accreted)
LMCs have binding energies lower than the observed LMC for a MW mass of $2
\times 10^{12}\,\msun$.  The energetics favor a combination of a massive MW halo
and a late-accreted LMC.

\subsection{Eccentricity}
\label{subsec:eccentricity}
\begin{figure}
 \centering
 \includegraphics[scale=0.55, viewport=0 0 410 440]{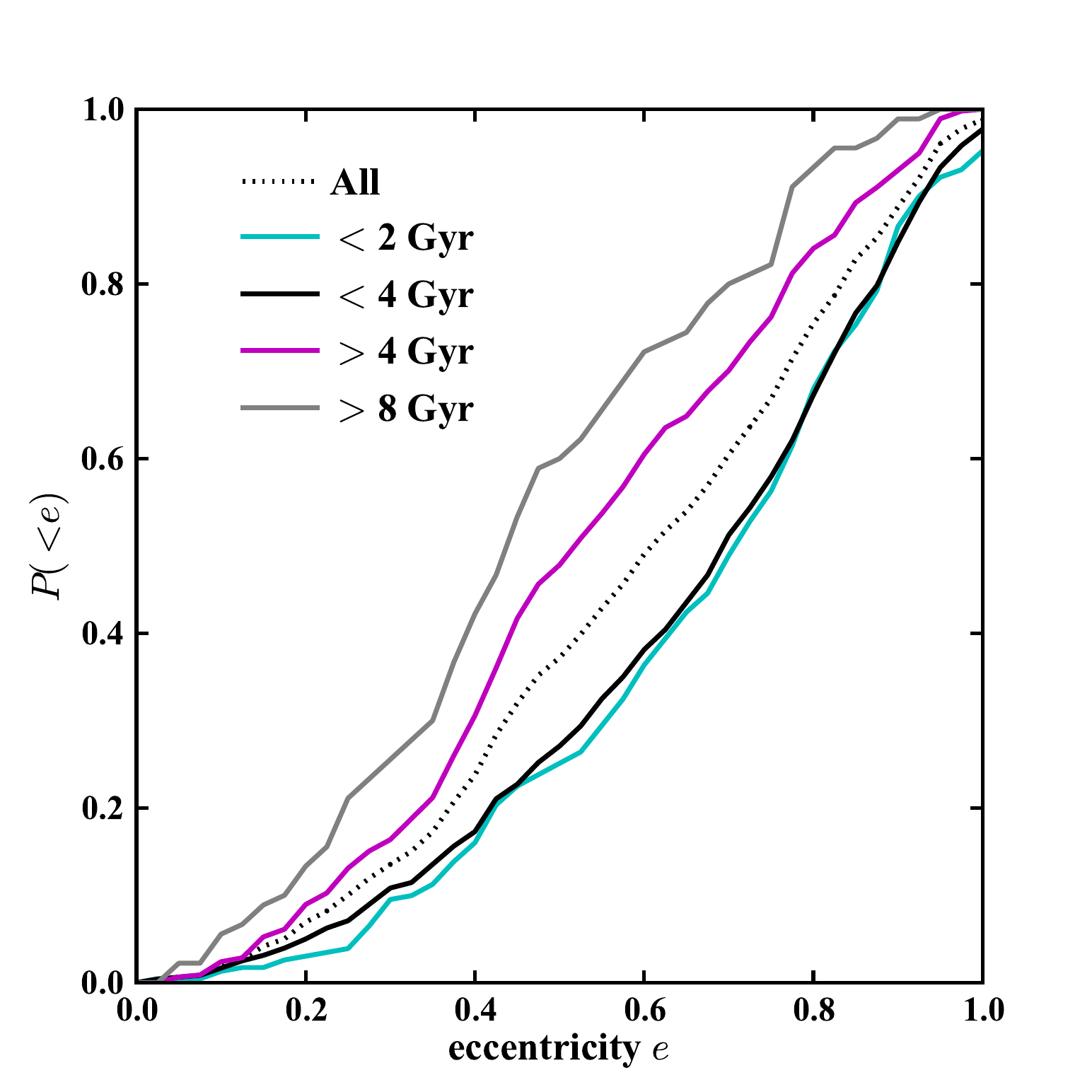}
 \caption{Orbital eccentricity for all LMC analogs (dotted line), as well as
   split by accretion epoch (solid lines).  Early-accreted LMCs (gray
   and magenta lines) have orbits that are significantly more circular than do
   late-accreted LMCs (cyan and black).  A MW of mass
   $(1.5,2,3)\times 
   10^{12}\,\msun$ corresponds to an LMC eccentricity of $(0.92, 0.79, 0.62)$, while a
   $10^{12}\,\msun$ MW results in an unbound orbit ($e > 1$).  
  \label{fig:eccentricity}
}
\end{figure}
To further explore the typical orbital properties of surviving LMC analogs as a
function of accretion epoch, we consider the distribution of orbital
eccentricities for these systems.  Orbital eccentricity $e$ is here defined as a
combination of the pericenters $r_p$ and apocenters $r_a$ of orbits:
\begin{equation}
  \label{eq:ecc}
  e \equiv \frac{r_a-r_p}{r_a+r_p}\,.
\end{equation}
(We assign unbound orbits an eccentricity larger than 1.)  With this definition,
$e < 0.5$ corresponds to fairly circular orbits ($r_a/r_p < 3$),  while $e=0$
indicates a perfectly circular orbit.

Figure~\ref{fig:eccentricity} shows that early-accreted LMC analogs tend to be
on orbits that are substantially more circular than those of late-accreted LMC
analogs.  Only 20\% of LMCs accreted within the last 2 or 4 Gyr (cyan and black
solid lines, respectively) have $e< 0.5$, a value that is met by approximately
50\% of LMCs accreted over 4 Gyr ago and by 60\% accreted more than 8 Gyr in the
past.  Although late accretion does not {\it a priori} mean that the Clouds
cannot have completed multiple pericentric passages, these data on
eccentricities provide further constraints.  The mean eccentricity for the
late-accreted LMC analogs is $\sim$ 0.7.  Given that the pericenter of the LMC's
orbit is $\sim$ 45 kpc, the resulting apocenter is $\sim$ 260 kpc.  The MCs
clearly could not have completed multiple pericentric passages on such an orbit
within the past 4 Gyr.  We therefore conclude that the MCs were most likely
accreted within the past 4 Gyr and are on their first passage about the MW

Fig.~\ref{fig:eccentricity} reinforces how unlikely it is to find an object with
the LMC's infall mass and orbit in a $10^{12}\,\msun$ halo at $z=0$, regardless
of accretion epoch, as
the LMC has $e >1$ according to our definition -- i.e., the LMC's orbit is
unbound -- for this mass.
If the MW {\it does} have such a low-mass halo, then the LMC was certainly
accreted within the past 4 Gyr.  Moreover, less than 10\% of orbits --
independent of accretion epoch -- have $e > 0.92$, which corresponds to a halo
mass of $<1.5 \times 10^{12} \,\msun$. It is therefore quite unlikely that the
Milky Way has a mass of less than $1.5 \times 10^{12}\,\msun$.  (For reference,
a Milky Way halo of mass $2 \times 10^{12}\,\msun$ results in an eccentricity of
0.79 for the LMC, while a mass of $3 \times 10^{12} \,\msun$ gives an
eccentricity of 0.62.)

\section{Frequency of LMC/SMC Analogs about MW Hosts}
\label{section:Mass}
\begin{figure*}
 \centering
\includegraphics[scale=0.65, viewport=10 0 810 585]{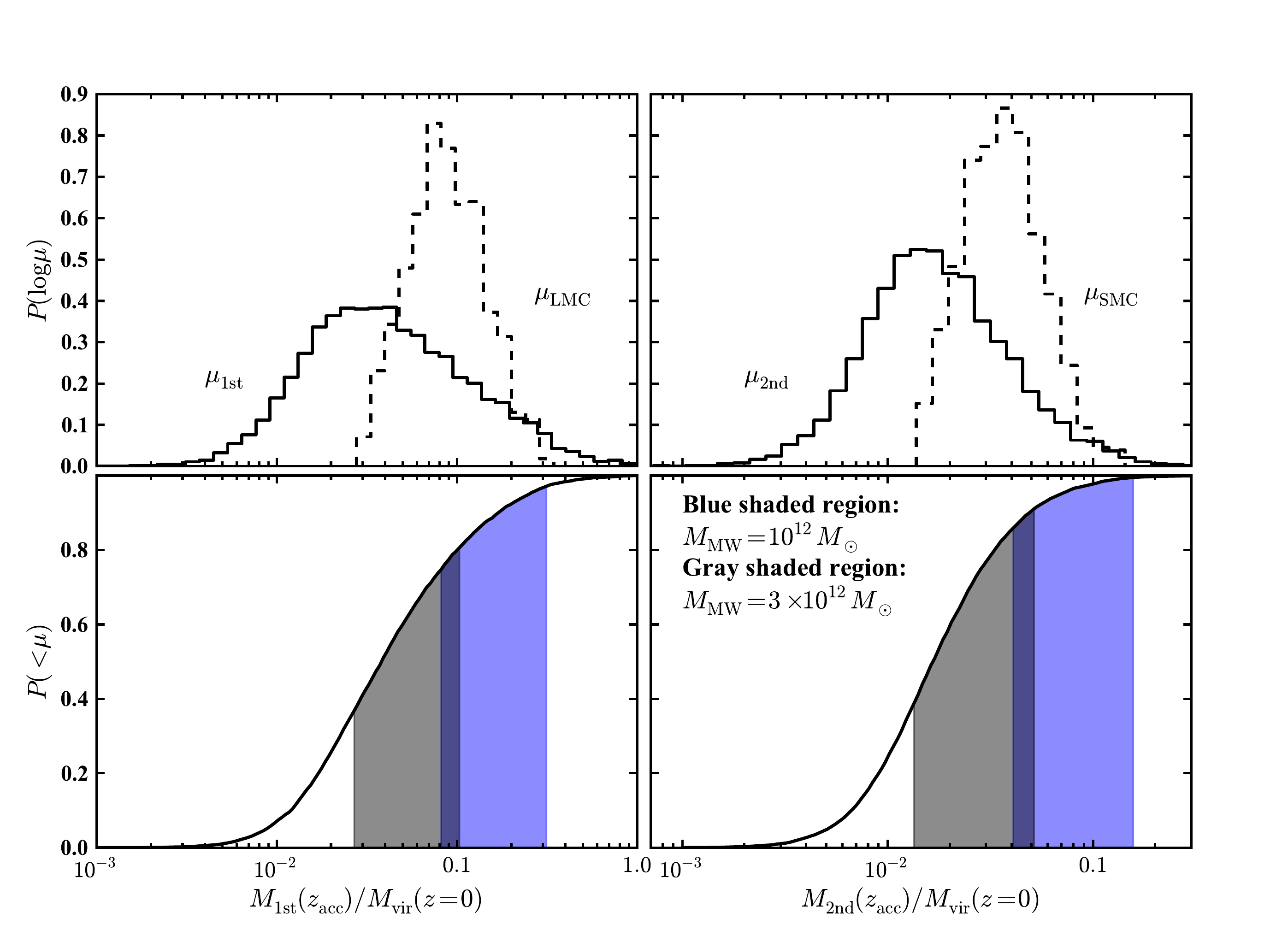}
 \caption{ The distribution of masses of first-ranked subhalos (left panels) and
   second-ranked subhalos (right panels) with respect to the virial mass of
   their hosts.  {\it Upper panels:}
   Distribution of $M_{\rm 1st} /\mvir(z=0)$ and $M_{\rm LMC}/\mvir(z=0)$
   (solid and dashed histograms, left panel), and of $M_{\rm 2nd} /\mvir(z=0)$
   and $M_{\rm SMC}/\mvir(z=0)$ (solid and dashed histograms, right
   panel). {\it Lower panels:} Cumulative distributions of solid curves from
   upper panels.  The shaded regions correspond to the observed range of infall
   masses for the LMC (left panel) and SMC (right panel) assuming a MW mass of
   either $10^{12}\,\msun$ (blue region) or $3\times 10^{12}\,\msun$ (gray
   region).     \label{fig:mu_cum}
}
\end{figure*}

In this section, we examine the mass distribution of the samples defined in
$\S$~\ref{subsubsec:MC_sim}.  Our goal is to determine how typical the infall
masses of the MCs are relative to the full first- and second-ranked subhalo
samples and as a function of the host mass.
 
The upper left panel of Fig.~\ref{fig:mu_cum} shows the distribution of masses
for first-ranked subhalos (solid histogram) and LMC analogs (dashed histogram),
relative to the redshift zero virial mass of their hosts.  The distribution of
masses for first-ranked subhalos is fairly broad and peaks at $M_{\rm 1st}
\approx 0.03 \mvir(z=0)$.  The mass distribution of LMC analogs is substantially
narrower and peaks at a much higher mass, $M_{\rm LMC} \approx 0.1 \mvir(z=0)$.
The difference between the two distributions indicates that the LMC is more
massive than the typical first-ranked satellite galaxy in a Milky Way-mass halo,
a conclusion also reached in BK10.

A similar situation exists for the SMC, which is shown in the upper right panel
of Fig.~\ref{fig:mu_cum}.  While the distribution of $\mu_{\rm 2nd}$ peaks at
$\sim 0.01$, the distribution of $\mu_{\rm SMC}$ peaks at $\sim 0.03$.  This
shows that the SMC is also more massive than the second-ranked galaxy in a
typical MW-mass halo.

The solid line in the lower left panel of Fig.~\ref{fig:mu_cum} shows the
cumulative version of $M_{\rm 1st}(\zacc)/\mvir(z=0)$.  The combined shaded
region encompasses the full LMC analog sample; the range of $\macc/\mvir(z=0)$
corresponding to $\mvir=10^{12}\,\msun$ is shown in blue, while the range for
$3\times 10^{12} \,\msun$ in shown in gray.  The lower right panel of
Fig.~\ref{fig:mu_cum} plots the same quantities for our sample of second-ranked
subhalos and SMC analogs.

Recall that there are 2658 host halos in the mass range $\mvir \in [1-3] \times
10^{12}\,\msun$.  If we define MC analogs strictly in terms of mass, we thus
conclude that approximately 35\% of MW-mass halos host an LMC analog and 32\%
host an SMC analog within $\rvir$.  These numbers are for a specific range of MW
halo masses, however, and are sensitive to the precise mass of the MW (BK10).  A
halo of $10^{12}\,\msun$ has a $\approx 20\%$ chance of hosting an LMC analog,
and less than a $10\%$ chance to host an SMC analog.  A $3 \times
10^{12}\,\msun$ MW makes L/SMC analogs much more common, as approximately $40\%$
of such hosts have L/SMC analogs.  [Note that these numbers are likely to
  be upper estimates, as we have used very conservative estimates on the errors
  for the mapping between $\mstar$ and $\macc$.]

In a search for LMC analogs in the seventh data release of the Sloan Digital Sky
Survey \citep{york2000, abazajian2009}, Tollerud et al. (in prep.) determine
that approximately 40\% of isolated hosts with luminosities similar to that of
the MW also have an LMC analog (r-band magnitude between -17.5 and -20) located
within a projected separation of 250 kpc.  Although the selection criteria of
Tollerud et al. differ from those used here, this result appears consistent with
our findings.

\begin{figure}
 \centering
 \includegraphics[scale=0.55, viewport=0 0 410 440]{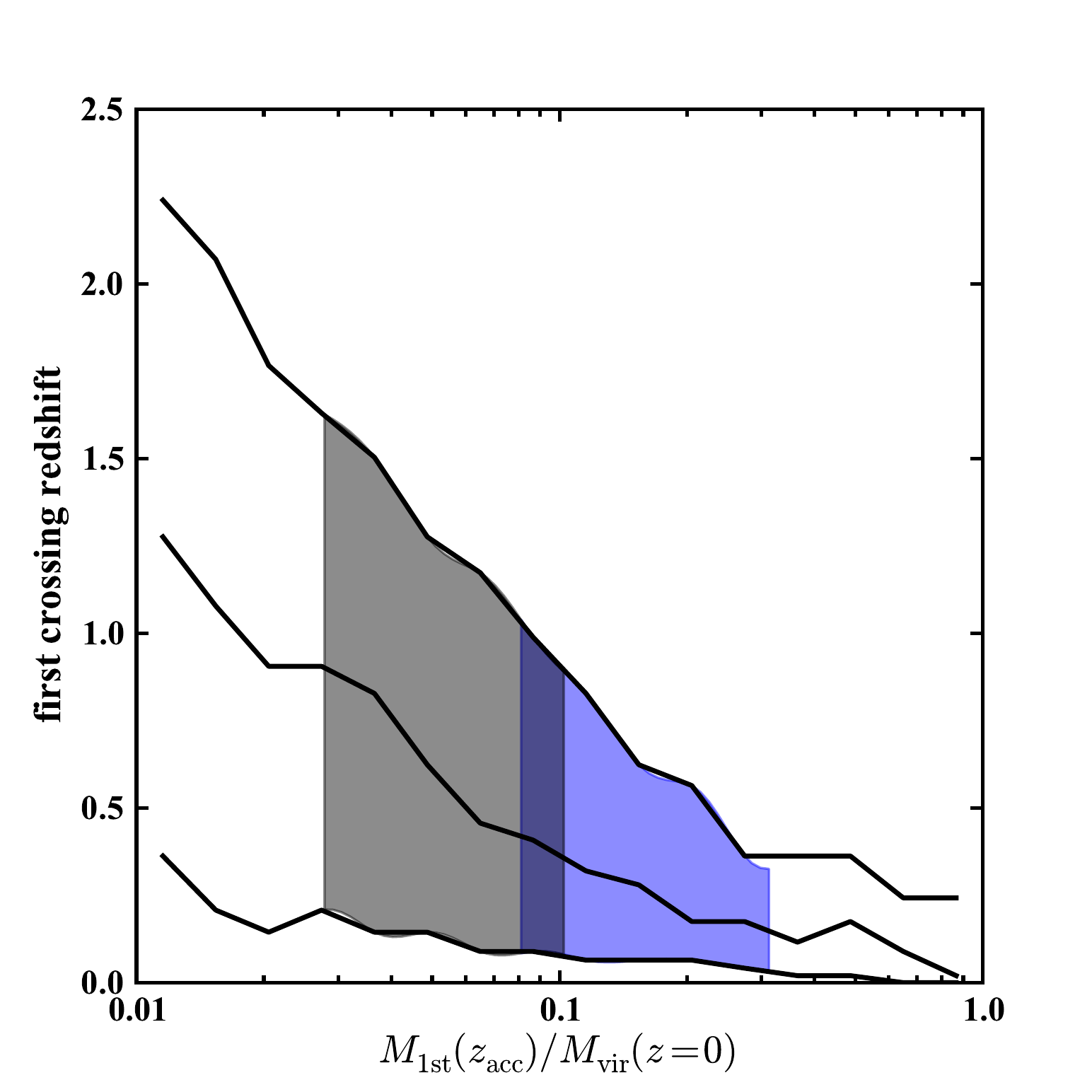}
 \caption{Redshift $\zfc$ where the first-ranked subhalo 
   first crossed the $z=0$ virial radius of its host 
  as a function of $M_{\rm 1st}/\mvir(z=0)$.  The middle line shows the
   median, while 70\% of the  distribution is contained between the upper and
   lower lines. The shaded region(s) are the same as in Fig.~\ref{fig:mu_cum}.
 \label{fig:firstCross}
}
\end{figure}

The mass of the first-ranked subhalo correlates strongly with its first crossing
redshift $\zfc$, which is shown in Fig.~\ref{fig:firstCross}.  The median
(middle curve) decreases from $\zfc \approx 1.4$ at $\mu=0.01$ to $0.4$ at
$\mu=0.1$ (70\% of the distribution is contained within the dashed curves).  The
shaded region in Fig.~\ref{fig:firstCross} shows the allowed range for the LMC
when using the abundance matching assumption (see Fig.~\ref{fig:mu_cum})

\begin{figure}
 \centering
 \includegraphics[scale=0.55, viewport=0 0 410 440]{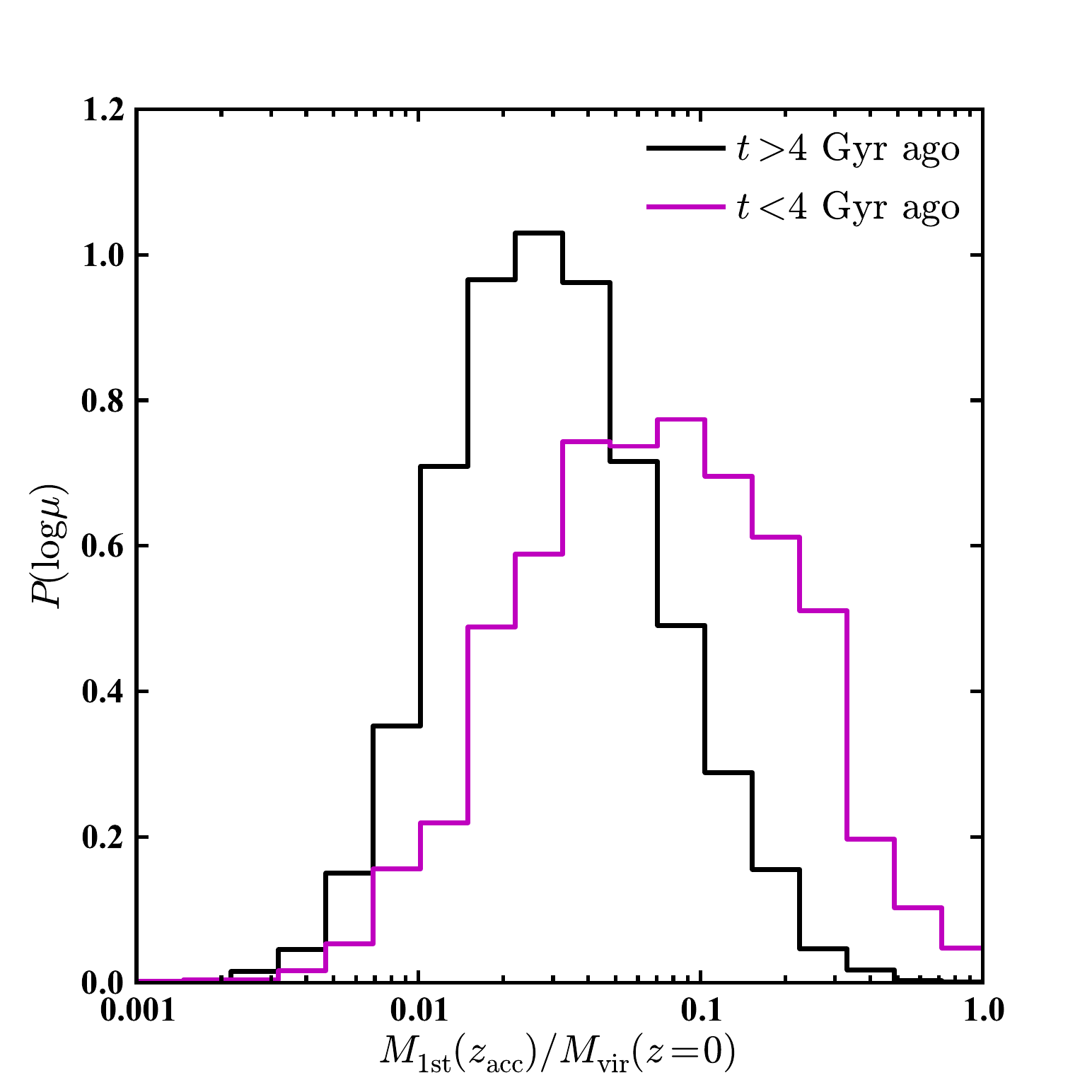}
 \caption{Distribution of masses for 1st ranked subhalos accreted early
   (more than 4 Gyr ago; black) and late (within the last 4 Gyr; magenta).  Note
   that each 
   distribution is separately normalized to unity; the number of first-ranked
   satellites with early accretion times is approximately 2.4 times greater
   than those with late accretion times. Those accreted at late times tend 
   to be a factor of $\sim$3 more massive at infall than those accreted at 
   early times. 
  \label{fig:mHist}
}
\end{figure}

An alternate way of looking at the dependence of $\mu({\rm 1st})$ on $\zfc$ is
to divide the 1st ranked subhalos into two samples: those accreted early,
defined here as $\zfc > 0.4$ (or equivalently, a look-back time of $t > 4$ Gyr,
corresponding to the local minimum in Fig.~\ref{fig:firstCrossDist}), and those
accreted at late times ($\zfc < 0.4$ or $t < 4$ Gyr).  The result of this split
is shown in Fig.~\ref{fig:mHist} and reinforces the result of
Fig.~\ref{fig:firstCross}: 1st ranked subhalos that have joined their $z=0$ host
halos within the last 4 Gyr tend to be a factor of $\sim 3$ more massive at
infall than those that joined their host halo more than 4 Gyr ago.  Since the
LMC has been shown to be more massive than the typical 1st ranked halo (see
$\S$~\ref{section:Mass}), it is more likely to have been accreted at late times.

\begin{figure}
 \centering
 \includegraphics[scale=0.55, viewport=0 0 410 440]{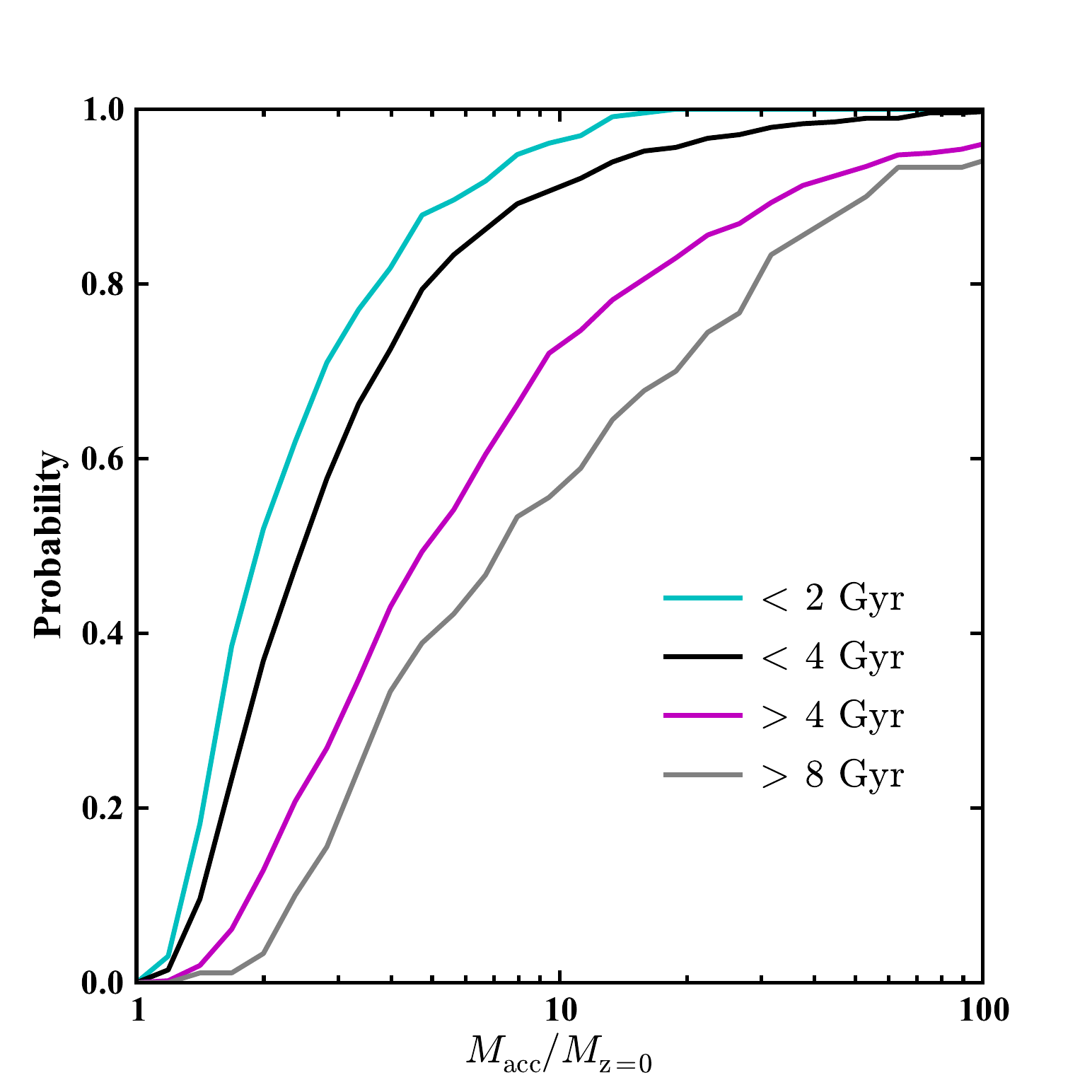} 
 \caption{The cumulative
   distributions of infall mass relative to $z=0$ mass (in dark
   matter) for LMC analogs.  Each curve corresponds to LMC analogs with
   different accretion epochs.  The most recently accreted candidates (cyan
   line) have typically lost of order half of their $\macc$, whereas the
   earliest accreted LMCs (gray line) have lost almost 90\% of their dark matter mass.
 \label{fig:sham_massLoss}
}
\end{figure}

In order to assess whether a late or early accretion scenario is more plausible
for the LMC, we need to also account for the survivability/mass loss expected
for LMC analogs, which will depend on the accretion epoch and the specific orbit
of the subhalo.  Figure~\ref{fig:sham_massLoss} shows the distribution of
accretion mass relative to redshift zero dark matter mass\footnote{This is the
  bound mass determined by the {\tt SUBFIND} algorithm.} for LMC analogs.
Results are plotted for early (more than 4 Gyr ago; cyan) and late (less than 4
Gyr ago; black) accretion epochs, as well as for very early (more than 8 Gyr
ago; gray) and very recent (less than 2 Gyr ago; cyan) accretion epochs.  There
is a pronounced, and not unexpected, trend for stronger mass loss in
earlier-accreted LMC analogs.  The most recently accreted LMCs are typically a
factor of 1.5-4 less massive at $z=0$ than at accretion (60\% confidence
interval), whereas the earliest accreted LMCs tend to be 3.5 to 30 times less
massive.  It is unlikely that the LMC could have undergone strong tidal
stripping without losing much of its gas, which is at odds with observations.

Massive satellites typically do not survive for long before merging with their
hosts: the dynamical friction timescale for a 1:10 object at $z=1$ is
approximately 5 Gyr \citep{boylan-kolchin2008}, shorter than the time between
$z=1$ and the present day.  This is corroborated by \citet{stewart2008}, who
showed that LMC-mass objects ($M \sim 10^{11}\, \msun h^{-1}$) typically do not
survive for more than $\sim$3 Gyr after accretion (their figure 5).  BK10
examined the accretion epochs of massive subhalos in MW-mass hosts and found the
same trend (their Fig. 12).  Both groups argued, as we do here, that this lends
support to a first passage scenario for the LMC.

\section{LMC-SMC pairs}
\label{section:LS}
The small separation in position ($\sim$20 kpc) and velocity space ($\sim$100
km/s) between the two Magellanic Clouds makes it unlikely for them to be
coincidental neighbors.  Furthermore, the existence of (1) a bridge of gas
connecting the two galaxies and (2) the Magellanic Stream, a stream of HI gas
that trails behind the MCs over 150$^{\circ}$ across the sky, also strongly
suggests that they have interacted for at least some time in the past
\citep{besla2010}.

These observations lead to a number of questions regarding the MC system,
including: What is the probability that the MCs were accreted together and
survive as a binary today?  Is there a preferred accretion epoch or mass ratio
for such a pair?  How likely is it that the Clouds are only an apparent binary
system today rather than a true binary system?

\begin{figure}
  \centering
  \includegraphics[scale=0.55, viewport=0 0 410 440]{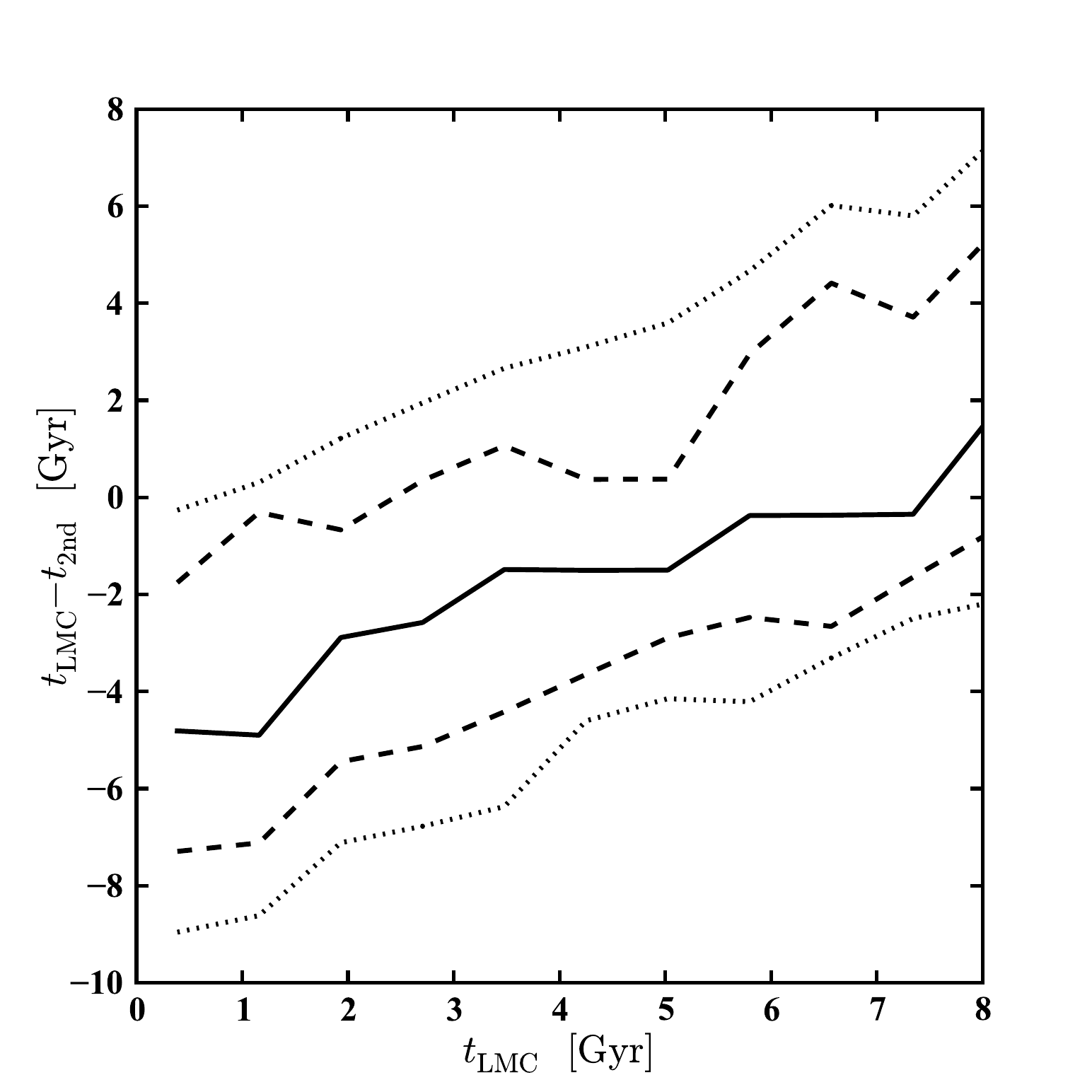}
  \caption{Difference in first crossing times between LMC analogs and the second
    ranked subhalo in each
    system as a function of the first crossing time of the LMC.  (All times are
    lookback times.)  The dotted (dashed) lines show the 10-90 (25-75)
    percentiles of the distribution, while the solid line shows the median.  For
    LMCs with $\tfc \sim 4$ Gyr, almost 40\% of SMCs were accreted within $\pm 2$
   Gyr of the LMC, and 25\% within $\pm 1$ Gyr.  
   \label{fig:pairs_tdiff}
 }
\end{figure}

To this end, we consider the difference in accretion epoch between each
LMC analog and the corresponding second-ranked halo about the same host.
Figure~\ref{fig:pairs_tdiff} shows the median difference (solid line) in $\tfc$
for the LMC analogs and second-ranked subhalos as a function of $\tfc$ for the
LMC analog, as well as the 10\% (dotted) and 25\% (dashed) quantiles.  It is
indeed possible to find first and second-ranked subhalos that are accreted as
pairs (within 1 Gyr of each other).  This probability depends on the LMC
accretion epoch, however, and does not provide information about whether the
accreted pairs can remain as a binary to the present day.  The number of pairs
accreted $>$ 4 Gyr is approximately three times larger than the number accreted
within the last 4 Gyr.  This is likely a result of 1 Gyr being a much larger
fraction of a typical orbital time at higher redshifts than today.

We can also compute how likely it is to find MC-like objects around MW-mass
hosts in a binary system at $z=0$.  To do so, we calculate the separation in
position and velocity space between the LMC analog sample and the corresponding
second-ranked subhalo for the same host.  Systems having $|\Delta v| < 150
\,{\rm km/s}$ and $|\Delta R| < 50 \,{\rm kpc/h}$ are considered plausible
binaries.  Table 1 lists a number of properties for the 23 identified LMC/SMC
analogs, including the first crossing time for each Cloud. Recall that the LMC
analog sample contains 938 halos; it is thus possible, though not probable
($\sim 2.5\%$), that an LMC analog and a second ranked halo be found in a binary
system about a MW-mass host today.  

Since SMC analogs represent a subset of the
second ranked subhalo sample, these numbers also indicate that LMC / SMC analogs
accreted at similar epochs are not likely to exist as binaries at the present
day.  The binaries that do survive to $z=0$ tend to have the SMC in an eccentric
orbit about the LMC.  This is in agreement with the proper motion measurements
of the SMC by \citet{kallivayalil2006a} and \citet{vieira2010}, who find a high
relative velocity between the Clouds. An eccentric SMC orbit about the LMC is
also required in the Magellanic Stream model proposed by \citet{besla2010}.  The
present study shows that such an orbital configuration is cosmologically
expected.

\begin{table}
  \caption{
    Properties of LMC-SMC binaries.  Column 1: ratio of masses of most massive
    progenitors; column 2: mass ratio at $z=0$; column 3: first crossing time
    for LMC; column 4: first crossing time for SMC; column 5: separation between
    LMC and center of host; column 6: 3-D velocity of the LMC, relative to host.
  \label{table:binaries}
  }
  
\begin{tabular}{cccccc}
  \hline
  \hline
  $M_{\rm L}/M_{\rm S}$ & $M_{\rm L}/M_{\rm S}$ &$t_{\rm fc,LMC}$ & $t_{\rm fc,
    SMC}$ & $R_{\rm LMC}$ & $V_{\rm LMC}$\\  
  ~[$\zacc$]~& [$z=0$]$^a$ & ~[Gyr]~ & [Gyr] & [kpc] & [km/s]\\
  \hline
   11.74  &     20.01  &      0.00  &      0.26  &    244.39  &    121.92\\
    2.79  &     24.29  &      0.26  &      0.00  &    323.87  &    221.25\\
    6.42  &     16.27  &      0.54  &      0.26  &    180.25  &    225.66\\
    6.69  &     32.02  &      0.83  &      1.13  &    151.70  &    287.50\\
    3.20  &      1.30  &      1.44  &      1.13  &     26.09  &    417.31\\
   15.07  &     15.52  &      1.44  &      1.44  &     82.97  &     57.49\\
    7.76  &     16.18  &      1.76  &      1.13  &     98.58  &    127.10\\
   12.61  &     33.32  &      2.77  &      2.43  &    172.74  &     41.07\\
    1.11  &      2.60  &      2.77  &      7.56  &     88.77  &    124.46\\
    1.15  &      1.83  &      3.13  &      3.49  &    124.35  &    153.10\\
    1.35  &      2.25  &      3.49  &      5.73  &     41.78  &     99.13\\
    7.19  &     15.53  &      4.23  &      4.97  &    115.93  &    121.54\\
   21.24  &    151.66  &      4.97  &      4.60  &    148.06  &    129.51\\
    1.39  &     21.76  &      5.35  &      7.91  &     95.29  &    132.18\\
    2.57  &      3.43  &      5.73  &      4.97  &    232.48  &    175.78\\
    1.86  &     13.26  &      6.47  &      5.73  &    104.46  &    163.18\\
    3.41  &     12.83  &      6.47  &     10.06  &     39.37  &    240.08\\
    5.27  &      9.49  &      6.84  &      7.20  &    107.45  &     92.04\\
    1.40  &      7.25  &      7.20  &      9.20  &    141.99  &     82.45\\
    1.77  &      0.83  &      7.20  &     10.06  &     57.44  &     22.06\\
   11.47  &     25.43  &      7.56  &     10.06  &     72.74  &    127.48\\
    1.45  &      0.73  &      7.56  &      9.20  &     94.08  &     25.78\\
    3.83  &      3.32  &      8.58  &      7.91  &    107.39  &    229.95\\
  \hline
\end{tabular}
$^a${\it Note that the calculation of subhalo masses at redshift zero is sensitive to the
  location of the subhalos within their hosts.}
\end{table}

A few specific cases from Table~\ref{table:binaries} stand out in particular.
The first three rows have the pairs that were accreted most recently.  All three
of these systems have more angular momentum than the true LMC, highlighting that
there may be no issue with the large magnitude of the LMC's angular momentum if
it was accreted recently.  Additionally, all are on fairly energetic orbits:
with $E/E_{\rm vir} \approx 1$, these are the most energetic among all 23 binary
candidates.  In fact, most of the other candidates have $E/E_{\rm vir} \ga 2$,
which places them on orbits that are improbably bound (relative to the observed
LMC) even for a MW of $10^{12},\,\msun$ (see Fig.~\ref{fig:e_cum}).  With the
exception of the oldest binary candidate (the final row of
Table~\ref{table:binaries}), all systems with $\tfc({\rm LMC}) > 5$ Gyr have low
angular momentum, likely due to losses via dynamical friction.  Although the
volume of the MS-II does not provide a vast sample of possible binaries, those
that do exist with orbital energetics similar to that observed for the Clouds
are accreted at late times. Specifically, the systems highlighted in the first
three rows of Table 1 illustrate that it is possible for a binary LMC/SMC to be
accreted very recently on a high angular momentum/energy orbit

A further point of interest is that the pairs with first crossing redshifts that
differ by $\ga 2.5$ Gyr between the LMC and the corresponding second ranked
subhalo are all cases where the LMC is within 100 kpc of the host.  If these
systems were true binaries, then there should also exist examples where the two
subhalos had discrepant accretion epochs and are located at large distances from
the host today. Since no such examples exist, it is likely that these are chance
associations of two satellites near pericenter than true binaries.

\section{Discussion}
\label{section:Discuss}
Using the Millennium-II Simulation, we have investigated \lcdm\ predictions for
orbital properties and accretion histories of the MCs in the context of the
updated proper motion measurements of Kallivayalil et al. (2006a,b).  In this
section, we further explore how environment or cosmological parameters may
influence our results, the likelihood that the MCs are on their first passage
about the MW, and the masses of halos hosting subhalos with LMC-like separations
and velocities.

\subsection{Milky Way sample}
\begin{figure}
 \centering
 \includegraphics[scale=0.55, viewport=0 0 410 440]{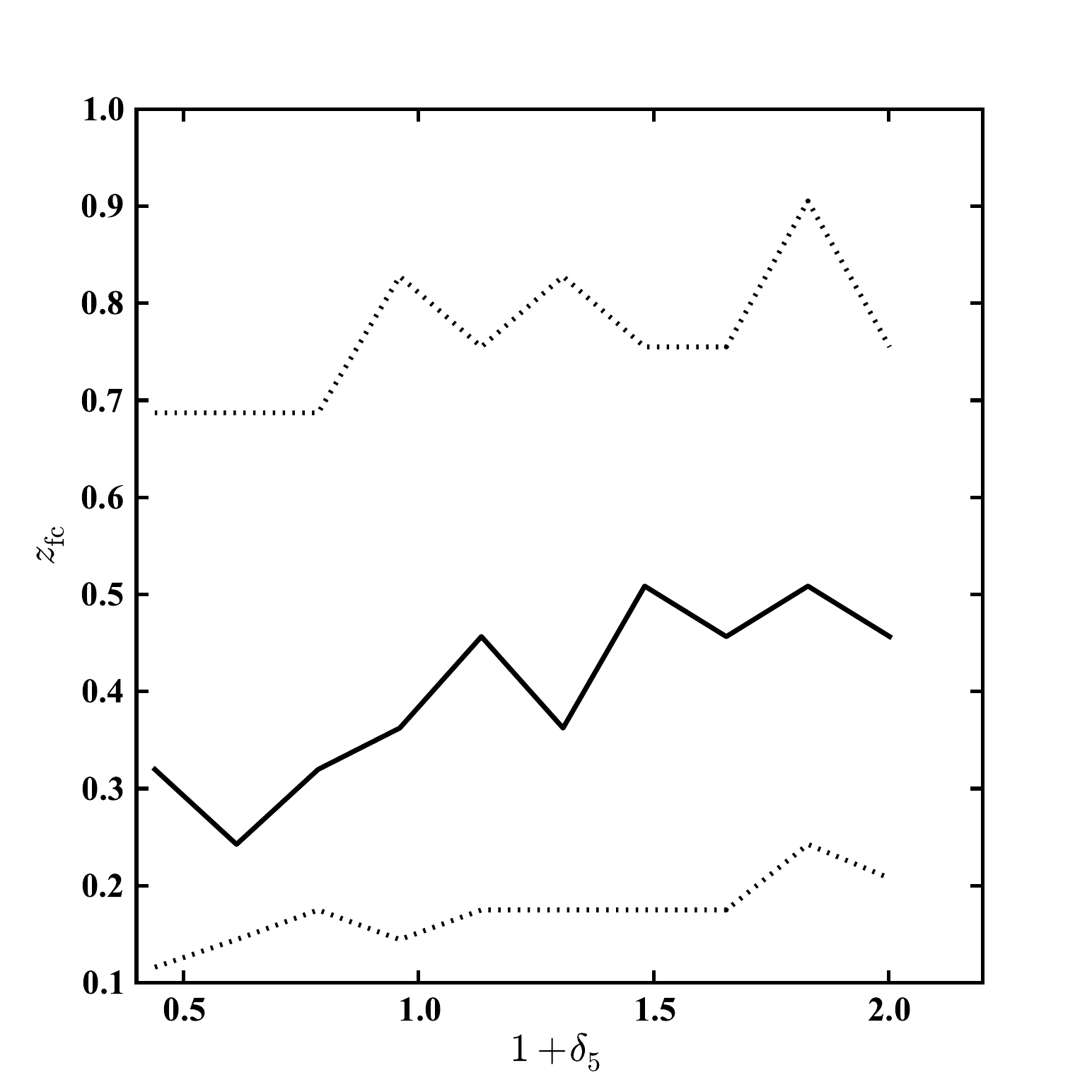}
 \caption{Dependence of $\zfc$ on the large-scale overdensity, measured via
   gaussian smoothing on $5\,\hmpc$ spheres, for the LMC analogs.  There is no
   obvious dependence of $\zfc$ on the large-scale overdensity, indicating that
   the large-scale environments of MW-mass hosts does not strongly influence the
   accretion epochs of their most massive subhalos.
  \label{fig:zfc_dens}
}
\end{figure}

It is important to investigate whether the trends shown in the previous sections
have any systematic dependence on any properties of the host halos.  In
particular, halos residing in low density environments have different accretion
histories than those in high density regions \citep{gottlober2001,
  maulbetsch2007, fakhouri2010}, an effect that could
potentially bias our results on accretion epochs of LMC analogs.  We therefore
plot how the large-scale environment of a halo influences the first crossing
redshift of LMC analogs in Fig.~\ref{fig:zfc_dens}.  This plot shows that $\zfc$ is
essentially independent of environment as measured by the dark matter
overdensity, smoothed with a Gaussian filter of width $5\,\hmpc$.  Our results
should therefore be insensitive to the environment of the host halo.

The typical accretion epoch of LMCs could also be affected by the choice of
cosmology in the \millen, which has $\sigma_8$ that is approximately 10\% higher
than the current best-fitting value of $\approx 0.81$ \citep{komatsu2009}.
[This difference is minor in terms of the number of MW-mass halos found at
$z=0$, as it affects the abundances of such halos by 10\%.]  While a
quantitative understanding of the effects of varying $\sigma_8$ would require
running a completely new simulation, we can easily estimate the qualitative
effect by noting that lowering $\sigma_8$ results in later formation of halos of
a given mass.  We therefore expect that any changes due to reducing $\sigma_8$
would tend towards even later accretion epochs for LMC (and SMC) analogs.

Alternatively, we can note that $\sigma_8(z=0.21) \approx 0.81$ for the
\millen\ cosmology; the difference between the distribution of $\tfc$ for halos
defined at this epoch and at $z=0$ should therefore inform us about potential
differences due to changes in the power spectrum normalization.
\begin{figure}
 \centering
 \includegraphics[scale=0.55, viewport=0 0 410 440]{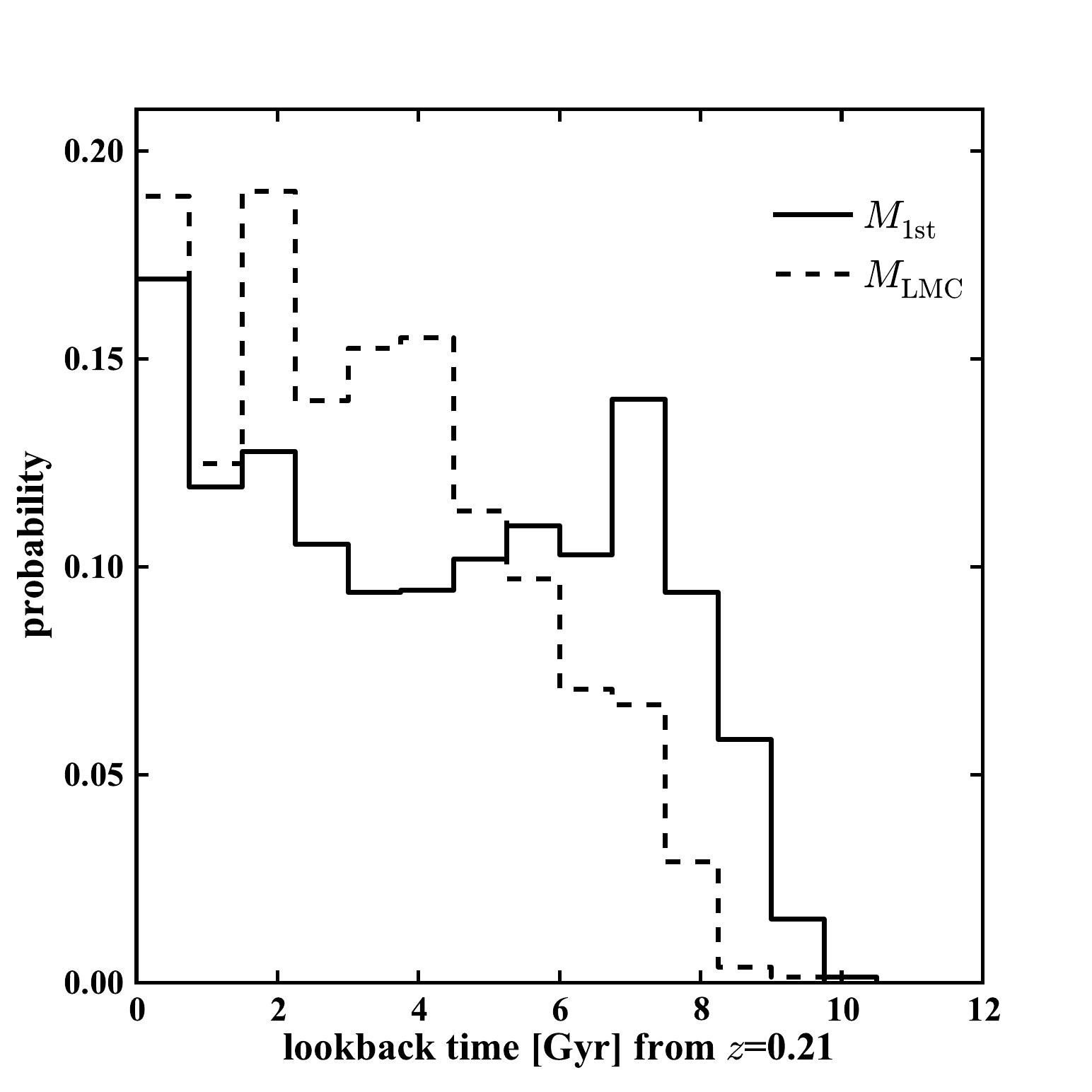}
 \caption{Distribution of first crossing times for the sample of MW-mass hosts
   selected at $z=0.21$ (analogous to upper panel of
   Fig.~\ref{fig:firstCrossDist}).  This distribution is similar to the
   distribution for $z=0$ hosts, suggesting that changing $\sigma_8$ from 0.9 to
   0.815 should not affect our conclusions on the probability of the LMC being
   on its first pass about the Milky Way.
  \label{fig:zfc_z02}
}
\end{figure}
The distribution of first crossing times for 1st-ranked subhalos and LMC analogs
of halos selected from the \millen\ at $z=0.21$ (using the same criteria
described in \S~\ref{section:methods}) is shown in Fig.~\ref{fig:zfc_z02}.  Comparing this
distribution with the distribution of $\tfc$ in halos selected at $z=0$ (the upper
panel of Fig.~\ref{fig:firstCrossDist}), we can see that little changes between
$z=0$ and $z=0.21$.  Accordingly, reducing $\sigma_8$ from the Millennium and
\millen\ value of 0.9 is not likely to strongly affect our findings on the
accretion epochs of LMC-like satellites.

\subsection{Are the Magellanic Clouds on their first passage 
  about the Milky Way?} 
\label{subsection:First}
A number of lines of observational evidence support the idea that the Magellanic
Clouds are making their first pericentric pass about the Milky Way.  This
scenario explains why two gas-rich satellites reside at small galactocentric
distances -- similar satellites are typically found at much larger distances
from the MW or M31 \citep{van-den-bergh2006} -- as the Clouds would not have had
sufficient interaction with the MW to have lost their gas by some combination of
tidal stripping, harassment, and ram pressure stripping \citep{mayer2006}.
Similarly, the unusually blue color of the LMC (\citealt{james2010}; Tollerud et
al., in preparation) means that it must have retained a substantial amount of
star-forming gas, which is difficult to understand if the MCs have completed
multiple orbits about the MW.  The existence of stellar populations extending as
far as $\approx 20$ kpc from the LMC's dynamical center \citep{munoz2006,
  majewski2009, saha2010} is also an indication that the LMC has not interacted
strongly with the MW.  Finally, \citet{besla2010} have shown that the Magellanic
Stream may originate from a tidal interaction between the MCs themselves, a
model that requires the MCs to be a recently-accreted binary system.

B07 first examined the possibility that the MCs have been recently accreted by
the MW using an orbital analysis constrained by the new HST proper motion
measurements.  Uncertainties in modeling the MW meant that a scenario in which
the Clouds were accreted at early times cannot be ruled out by such an analysis
(e.g., \citealt{gardiner1996, besla2007, shattow2009}), and refinements in the
error space of the proper motions are unlikely to improve this situation.  (We
note, however, that early accretion models assume a static MW halo potential
over a Hubble time, which is unrealistic.)  We have computed the likelihood of a
first passage scenario using a large sample of MW-mass halos from a high
resolution cosmological simulation of the \lcdm\ cosmology.  Our results put the
first passage scenario on even firmer ground.  We have showed that it is highly
improbable for surviving satellites with infall masses similar to that of the
LMC to have been accreted at $z>1$, rendering an early infall scenario unlikely
based on mass considerations alone.  We further found that 25\% of surviving LMC
analogs have been accreted within the past 2 Gyr and that the energetics of the
LMC orbit are strongly inconsistent with the properties of LMC analogs accreted
$>$ 4 Gyr ago.  Finally, we have showed that recently-accreted LMCs are
incapable of making multiple pericentric passages by the present. Taken
together, these results demonstrate that it is quite likely that the MCs have
recently joined the MW and are currently making their first pericentric pass; a
{\it very} recent accretion ($\la$ 2 Gyr ago) is also cosmologically plausible.

\subsection{Mass of the Milky Way}
\label{subsec:bayes}
Uncertainties in the mass of the Milky Way's halo play a substantial role in
placing the LMC and its orbit in a cosmological context, as has been shown
repeatedly in our analysis.  A low mass halo ($\mvir \approx 10^{12}\,\msun$)
means that the LMC is fairly unusual in terms of its (high) mass and that it is
{\it very} unusual in terms of its (energetic) orbit.  Both the mass and orbital
energy of the LMC are more typical for halos of $\mvir \ga 2 \times
10^{12}\,\msun$.

\begin{figure}
 \centering
 \includegraphics[scale=0.55, viewport=0 0 410 440]{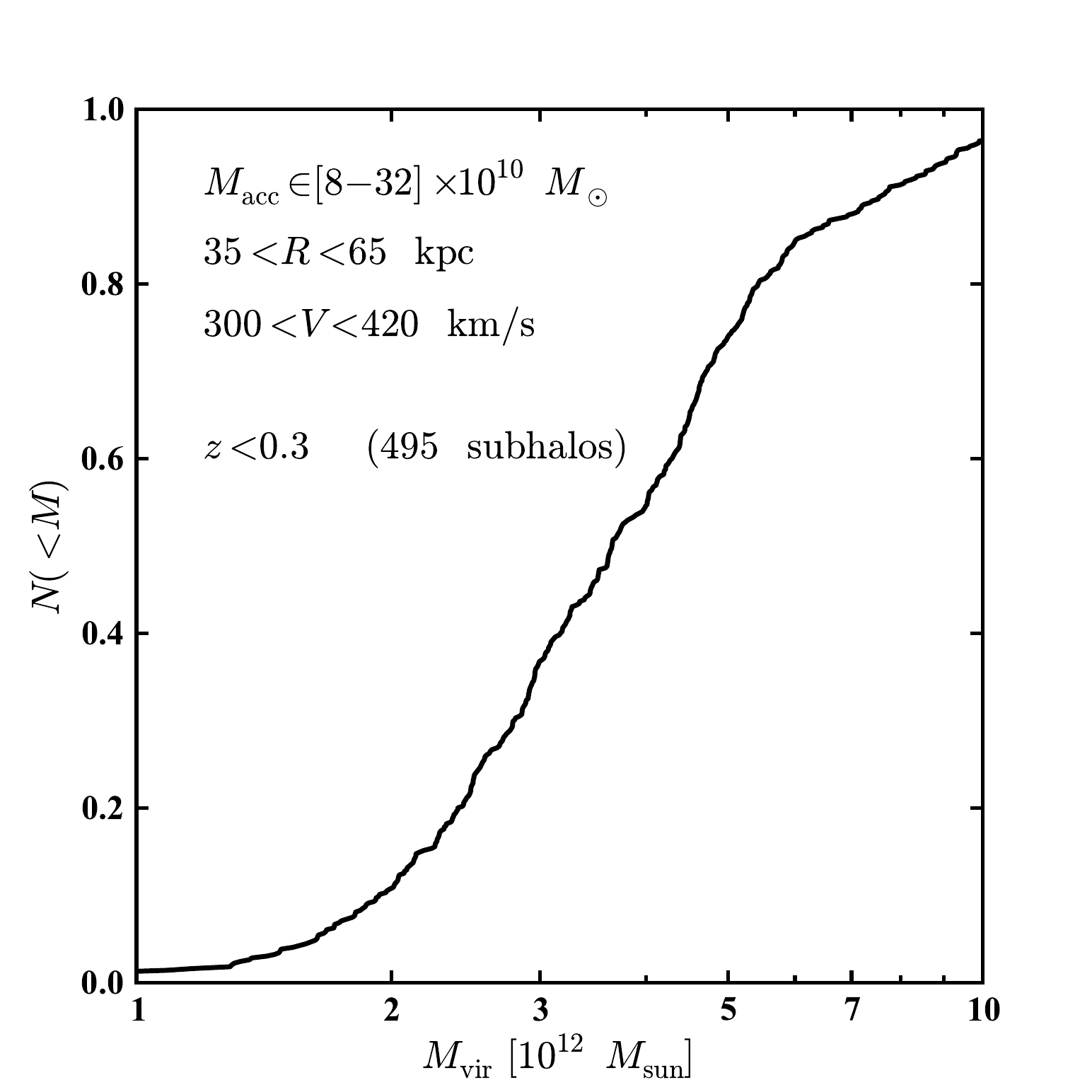}
 \caption{Distribution of halo masses in the \millen\ having a satellite similar
   to the LMC: a mass of $8 \times 10^{10}< \macc/ \msun < 3.2 \times
 10^{11}\,\msun$, at a separation of $35 < R < 65 \,\kpc$ from its host, and
 with a total velocity of $300 < V < 420 \,\kms$.  Of all the hosts with such a
 subhalo at $z<0.3$, over 90\% have $\mvir > 2\times 10^{12}\,\msun$.
  \label{fig:bayes}
}
\end{figure}

We can also take an ``inverse'' view and ask, in what mass dark matter halo do
objects with masses, velocities, and halo-centric distances similar to the LMC
reside?  To this end, we build a sample of LMC analogs with no constraint on the
properties of the host halo (note that this differs from the mass-selected
samples used up to this point).  We merely require the following properties of
the subhalo: (1) $0.8 < \macc \; [10^{11}\,\msun] < 3.2$, (2) $35 < R < 65
\,\kpc$ from its host, and (3) $300 < V_{\rm tot} < 420 \,\kms$.  Since these
criteria are somewhat restrictive, we search for hosts in all 10 \millen\ snapshots with
$z<0.3$.  We find 495 subhalos matching our search criteria; the distribution of
host halo masses for these matches are shown in Fig.~\ref{fig:bayes}.  The
figure confirms that satellites with properties similar to those of the LMC are
unlikely to reside in host halos with $\mvir \la 1.5 \times
10^{12}\,\msun$.  If we further require that $\mvir < 3\times 10^{12}\,\msun$,
we still find that approximately 70\% of hosts of LMC-like subhalos reside in
hosts with $\mvir > 2\times 10^{12}\,\msun$.

\section{Conclusions}
\label{section:Conclusions}
The new HST proper motions for the MCs \citep{kallivayalil2006,
  kallivayalil2006a} have forced us to re-evaluate our understanding of their
orbital history about the MW. The canonical picture, wherein the MCs are on a
quasi-periodic, slowly decaying orbit around the MW, has been thrown out.  We
are left instead with two possibilities: (1) The MCs are on their first passage
about the MW (late accretion); or (2) the MCs joined the MW $>$ 8 Gyr ago and
are now on a highly eccentric orbit, having already completed at least one
passage about the MW since infall (early accretion).

In this work, we have addressed the likelihood of these two scenarios by using
the \millen\ to place the MCs in a cosmological context in terms of their
accretion epoch, orbital properties, and masses.  Our primary results can be
summarized as follows:

\begin{itemize}
\item 
  {\it LMC analogs are accreted preferentially at late times}: only 15\%
  have $\tfc > 7.5$ Gyr ($\zfc > 1$), while approximately 30\% have been
  accreted within the past 2 Gyr.  Such numbers favor the late accretion
  scenario for the LMC.

\item
  {\it The LMC's angular momentum is not anomalously high}: 30-35\% of all LMC
  analogs have specific angular momentum matching that of the real LMC if the
  mass of the MW lies between $[1-3]\times 10^{12}\,\msun$.  The angular
  momentum of the LMC is more typical of halos at the massive end of this range
  than of those at the low-mass end.

\item 
{\it It is exceedingly unlikely for the LMC to be on an unbound orbit}: If
  the mass of the MW is less than $2 \times 10^{12}\,\msun$, the LMC has an
  orbit that is more energetic than 90\% of comparable systems (adopting the
  mean radial and tangential velocities of K06).  40\% of MW systems with $\mvir
  \in [2-3]\times 10^{12}\,\msun$ have an LMC analog with orbital energy
  comparable to that of the LMC.  It is highly unlikely for LMC-like subhalos to
  be on unbound orbits, which is the case for a low-mass MW, and none of the
  early-accreted LMCs are on unbound orbits.  The conclusion that the LMC is in
  fact bound to a massive MW, and yet accreted recently, cautions against the
  use of backward orbital integration schemes to determine the orbital histories
  of satellites over cosmic time.

\item 
  {\it Energetically, it is difficult to accommodate a scenario where the
    MCs have made multiple pericentric passages}: LMCs accreted at early times
  are on mostly circular orbits, at odds with observations.  LMCs accreted
  recently have not had time to complete more than one pericentric passage.

\item 
 {\it LMC and SMC-mass objects are not particularly uncommon in MW-mass halos}:
  In a refinement of the results presented in \citet{boylan-kolchin2010}, we
  find that 20-32\% of MW-mass halos host an LMC analog and 10-25\% host an SMC
  analog.  These results are consistent with the analysis of LMC analogs
  about MW type hosts located in the SDSS DR7 by Tollerud et al. (in prep.).
  The MCs become less typical if the host halo is lower in mass. 

\item
 {\it It is possible, but not probable, to find LMC-SMC binaries at z=0}: 
  We find a small number of Milky Way-mass systems ($\sim 2.5\%$) with LMC
  analogs have apparent LMC-SMC binaries.  

\item 
  {\it Subhalos with properties similar to that of the LMC reside
    preferentially in massive host halos}: Out of {\it all} dark matter halos
  (without restriction on mass) hosting objects with masses, velocities, and
  separations similar to the LMC, only 10\% have $\mvir <2 \times
  10^{12}\,\msun$, and less than 5\% have $\mvir < 1.5 \times 10^{12}\,\msun$.
\end{itemize}

Overall, our results support a scenario in which the LMC is a recent addition
(in the last 4 Gyr) to a fairly massive ($\mvir \ga 1.5 \times 10^{12}\,\msun$)
Milky Way.

\section*{Acknowledgments}
MBK thanks Simon White for many helpful discussions, including the suggestion of
the ``inverse'' argument of \S\ref{subsec:bayes}, and for a careful reading of
an earlier version of this manuscript.  The Millennium-II Simulation databases
used in this paper and the web application providing online access to them were
constructed as part of the activities of the German Astrophysical Virtual
Observatory.  This work made extensive use of NASA's Astrophysics Data System
and of the astro-ph archive at arXiv.org.

\bibliography{draft}
\label{lastpage}
\end{document}